\documentclass[pre,superscriptaddress,notitlepage,nofootinbib, twocolumn,dvipsnames]{revtex4-2}
\pdfoutput=1
\usepackage[left=0.6 in, right=0.6 in, top=0.7 in, bottom = 0.7 in]{geometry}
\usepackage{hyperref}
\usepackage{graphicx}
\usepackage[normalem]{ulem}
\usepackage{amsmath,amssymb,mathtools,bm, physics}
\usepackage[colorinlistoftodos,prependcaption,textsize=tiny]{todonotes}
\usepackage{booktabs, hhline}\hypersetup{
    colorlinks=true,
    linkcolor=blue,
    filecolor=magenta,      	
    urlcolor=cyan,
}

\usepackage{soul}
\usepackage{xcolor}

\renewcommand{\r}{\vb r}

\newcommand{\p}{\vb p}
\renewcommand{\k}{\vb k}

\newcommand{\<}{\langle}
\renewcommand{\>}{\rangle}
\newcommand{\Zbar}{\langle Z \rangle}

\begin{document}
\title{ Ion stopping from bound and free electrons in plasmas: A channel-mixed RPA approach with average-atom orbitals } 
\author{Zachary A.~Johnson}
  \email{zjohnson@lanl.gov}
  \affiliation{Center for Nonlinear Studies, Los Alamos National Laboratory, Los Alamos, New Mexico 87545, USA}
\author{Patrick J. Adrian  }
    \affiliation{Los Alamos National Laboratory, Los Alamos, New Mexico 87545, USA}
    \affiliation{ Pacific Fusion Corporation, Fremont, CA, 94538, USA }
\author{Joshua A. Leveillee}
    \affiliation{Los Alamos National Laboratory, Los Alamos, New Mexico 87545, USA}
\author{Charles E. Starrett  }
    \affiliation{Los Alamos National Laboratory, Los Alamos, New Mexico 87545, USA}
\begin{abstract}
Ion stopping in partially ionized plasmas often receives roughly equal contributions from bound and free electrons, yet bound electron contributions are usually treated at a very coarse level, with the exception of costly time-dependent density functional theory (TD-DFT) simulations. At energies above the Bragg peak, where linear response methods are a good approximation, more accurate treatment is both feasible and critical for predictive modeling. We develop a channel-mixed random phase approximation (cmRPA) dielectric response function that combines Lindhard free response with explicit average-atom bound state transitions. Validating against ambient condition experiments and TD-DFT simulations, we demonstrate excellent agreement for proton stopping near and above the Bragg peak. The computational efficiency of our approach enables systematic exploration of extreme conditions across a wide range of densities and temperatures. We find non-trivial stopping effects from channel mixing between bound and free transitions via RPA, bound-bound stopping contributions at low velocity, and predict significant bound electron contributions to stopping in tungsten at peak compression conditions relevant to inertial confinement fusion experiments. Applying our model to the warm dense matter stopping experiment of Malko et al. (2022), we evaluate whether improved bound-state modeling could resolve the reported theory-experiment discrepancy, finding that within the constraints of linear response theory, bound stopping mismodeling is unlikely to be the source of the observed deficit.

\end{abstract}
\maketitle


\section{Introduction}
The energy transport in materials undergoing nuclear reactions in the form of either fusion or fission, as well as in experiments with energetic ion beams, is strongly dependent on the location and rate of ion energy deposition. For fast ions the largest source of stopping power, the rate of change of kinetic energy over distance, comes from the scattering of electrons and the excitation and ionization of bound states. The accurate and efficient model prediction of the full atomic electronic response to fast ions is crucial for understanding the impact of stopping power on high energy density phenomena, technologies, and experiments.

The modeling of the stopping of ions by electrons has received significant attention since Bethe's \cite{Bethe:1930ku} quantum calculation appropriate for particles above the electrons' typical velocity, and Bohr's classical calculation for slower ions \cite{Bohr01011913}. In modern theory these limits would be described as the strong binary scattering regime for low velocity ions and the dynamically screened linear response regime appropriate at high velocities. The linear response calculation is compelling in that it accurately predicts stopping power for low Z ions above the Bragg peak \cite{hentschel2023improving, White_2022,PhysRevB.91.014306, Kononov_2026_nonlinear} which is typically the relevant regime for the fast ions produced via fusion reactions. The fact that the linear response calculation is based on a general dielectric response function means it can represent arbitrarily complicated properties of the target material including scattering effects \cite{PhysRevB.1.2362}, local field corrections \cite{Faussurier2025}, discrete state transitions \cite{PhysRevB.25.6310}, and many-body perturbation theory effects \cite{PhysRevLett.133.026403, RevModPhys.74.601}. These neglect non-linear corrections higher-order in the ion's charge, $Z$, but can be approximately accounted for by analytic corrections to Bethe's formula, see \cite{Salvat2022} for a recent analysis. Due to the combination of this theory and the many experiments available at ambient conditions, the stopping power of ions in materials at near ambient conditions has undergone significant progress. This progress was compiled into the Stopping and Range of Ions in Matter (SRIM) code, \cite{ZIEGLER20101818} and International Atomic Energy Agency (IAEA) experimental database of stopping experiments\cite{IAEA} which serve as de facto standards for ambient condition stopping, despite some smaller lingering uncertainties \cite{WITTMAACK201657, MONTANARI2024165336}.

For conditions beyond ambient, particularly the warm dense matter (WDM) regime of plasma physics, characterized by partial ionization and degeneracy, there is substantially less experimental verification\cite{malko2022importance}.  Largely due to the transient nature of the material, this means that the construction of models analogous to SRIM using combined experimental and theoretical data is not yet possible for conditions significantly different than ambient.  However, recent experiments \cite{frenje2019experimental, Malko2022-ws, zylstra2015measurement} have begun probing this regime and show some ability to distinguish between models. In particular, explicit multi-atom time-dependent density-functional theory simulations (TD-DFT)  \cite{White_2022} shows better agreement with measurement \cite{Malko2022-ws} than many well-used stopping models.

These TD-DFT simulations represent the highest accuracy method routinely used, but each simulation is expensive and represents only a single point in a high dimensional space of possible plasma conditions, incident velocities and charges. To meet the practical needs for stopping across parameter space, the average atom (AA) is often used as a relatively inexpensive alternative model of a plasma that is comparable in accuracy to many-atom DFT everywhere except when significant chemical bonding is present. In particular, free-electron stopping for several materials in warm dense matter conditions were shown to have very good agreement between AA based models and TD-DFT simulations above the Bragg peak \cite{hentschel2023improving}, where the linear response approximation used in RPA based calculations is most accurate. 

For electronic stopping involving both free and bound electrons, however, the agreement between TD-DFT and AA models is less clear, and significant discrepancies are shown between these models in the recent charged particle review \cite{TCCW2} and in the experiments \cite{Malko2022-ws}. In the case of the review, this is likely due to the missing core contribution from the use of pseudopotentials common in plane wave TD-DFT simulations. In the case of the experiment the bound state stopping model was done with the CBC model \cite{Barriga-Carrasco_Casas_2013} which uses a simple Bethe-like stopping term derived from Hartree-Fock bound states. The CBC model has also been extended to use AA input, however it was found to typically overestimate stopping \cite{White_2022}.

Other models for bound stopping based on the AA use the known free electron linear response as an approximate local density form of the bound electron linear response, such as the LDA-RPA model \cite{faussurier2010equation, wang1998unified, mehlhorn2026enhancedrpaldamodelion}, and the SLPA model which additionally computes stopping per bound state \cite{barriga2022stopping}. These models can obtain reasonable stopping results because the underlying dielectric function conserves relevant sum rules. Nevertheless, representing bound electrons as plane waves is an uncontrolled approximation since the sum rules constrain only integrated spectral weight, not its distribution over frequency space.

Instead of treating the response of bound electrons approximately with the free response, or with a Bethe-like logarithm, one might achieve better agreement with simulation and experiment by treating the bound state transitions explicitly. In this paper, we compute the bound electron response via AA matrix elements representing bound excitation and ionization. We then mix these responses with the free Lindhard response using the RPA approximation to generate an \emph{ab initio} dielectric response that includes bound and free response, collective plasmon excitations, and approximately the mixing between these channels. This allows for a more consistent treatment of electronic stopping than has been achieved so far and represents essentially an all-electron extension of Lindhard based models such as in \cite{hentschel2023improving} analogous to the move from valence only to all-electron TD-DFT simulation. 

In Section~\ref{sec:SP_theory} we detail the theory of the linear response stopping power calculation accounting for bound-bound and bound-free transitions. In Section~\ref{sec:results} we present results of this stopping power model for a variety of conditions, including the Malko et al. \cite{Malko2022-ws} stopping power experiment and high energy density tungsten for double-shell inertial confinement fusion plasmas. Throughout the paper we use Hartree atomic units where $\hbar = e = m_e = a_B = 1$, though we frequently utilize cgs units in plots and applications.


\section{Linear Response Stopping Power with Bound Electrons} \label{sec:SP_theory}
The fundamental assumption in this work is that the electronic density response to the incident ion or beam of ions is small enough that the response is only linearly dependent on the ion potential---a limit that becomes increasingly exact for ions traveling faster than the mean electron velocities in the material. With the additional assumption that the ions are infinitely massive compared to the electron, the standard linear-response stopping power can be expressed for homogeneous target materials in terms of the dielectric function, $\epsilon$ \cite{Kremp2004-kr}
\begin{align}
    \frac{dE}{dx} = \frac{-2 Z_b^2}{\pi v_i^2} \int^\infty_0 \frac{dk}{k} \int^{kv_i}_0 d\omega  \omega \Im[\epsilon^{-1}(k,\omega)], \label{eq:SP}
\end{align}

where $Z_b$ is the incident ion or beam particle charge which has velocity $v_i$, with $k$, and $\omega$ representing the momenta and energy transferred to the plasma.

In this section we will work through how to construct this dielectric under the assumption that one has already computed a set of orbitals from an AA or other approximate Schrodinger equation solver in a spherically symmetric geometry. For us these orbitals were computed using the tartarus code \cite{gill2017tartarus} which allows stable and fast computations from ambient temperature and pressure systems to high energy density conditions.%

\subsection{Construction of the dielectric}
We construct a dielectric response function dependent on all bound and free electronic responses via a summation over the irreducible polarizations, $\chi^0$, of each relevant channel,
\begin{align}
    \epsilon(k, \omega) &= 1 - V_k \Big [ \chi^0_{bb}(k,\omega) + \chi^0_{bf}(k,\omega) + \chi^0_{ff}(k,\omega) \Big ] ,\label{eq:epsilon}
\end{align}
where $V_k=4\pi/k^2$ is the Coulomb interaction in atomic units, and the subscripts on $\chi^0$ denote contributions from bound-bound (\emph{bb}), bound-free / free-bound (\emph{bf}), and free-free (\emph{ff}) transitions. 
The inverse of this dielectric is also useful,
\begin{align}
    \epsilon^{-1}(k, \omega) &= \frac{\Re[\epsilon(k, \omega) ] - i \Im[\epsilon(k, \omega) ]}{|\epsilon(k, \omega) |^2},
\end{align}
where we see the channel dependence is non-separable, with the real and imaginary responses of all channels mixed together in the denominator. For example, the energy loss function (ELF) that goes into stopping in Eq.~\eqref{eq:SP} is,
\begin{align}
    \Im[\epsilon^{-1}]^{\rm cmRPA} &= \frac{V_k \Im[\chi^0_{bb}+ \chi^0_{bf} + \chi^0_{ff}]}{ |1 - V_k(\chi^0_{bb}+ \chi^0_{bf} + \chi^0_{ff})|^2 } \label{eq:ELF}
\end{align}
where the $k, \omega$ dependence is implicit. Since this method mixes bound and free transition channels together, we will call it the channel mixed RPA (cmRPA), which is opposed to the way that energy loss functions are typically constructed in a Chihara-style decomposition \cite{johnson2012thomson, barriga2022stopping}, which we will refer to as unmixed RPA (umRPA)\footnote{One can also consider a resummed, RPA-like treatment for the bound contributions, as in $\Im[\chi^0_{bf}]/|1 - V_k\chi^0_{bf}|^2$, but typically for applications like x-ray Thomson scattering this is not what is done.    }
\begin{align}
    \Im[\epsilon^{-1}]^{\rm umRPA} &= V_k \left( \frac{\Im[\chi^0_{ff}]}{ |1 - V_k\chi^0_{ff}|^2 }   + \Im[\chi^0_{bf}]  + \Im[\chi^0_{bb}]\right) . \label{eq:ELF_unmixed}
\end{align}
Since stopping power is an integrated quantity, we assume it will be sufficient to approximate the exact irreducible polarizability with the standard independent particle bare susceptibility formed by average atom states, approximating all many-body effects with the static Kohn-Sham potential. In momentum space, for a homogeneous plasma, this polarization in the independent particle approximation looks like:  
\begin{align}
    \chi^0(\k,\omega) &= g_s n_I \sum_{\alpha, \beta} \frac{f_\beta - f_\alpha}{w - E_\beta + E_\alpha  + i \epsilon } |\<\phi_\beta| e^{-i \k \r} | \phi_\alpha \>|^2 \label{eq:chi_0},
\end{align} 
where $\phi$ denotes both bound and free single independent particle orbitals obtained from average atom DFT, and $f_\alpha$ is the Fermi-Dirac distribution for state $\phi_\alpha$ with a solved chemical potential $\mu$. The prefactor $g_s=2$ counts the electron spin since the independent particle bare susceptibility is spin-independent. The additional factor of the ion density, $n_I$ in front is due to moving from a single center response to the full plasma response\cite{blenski1992linear}. We have written the state sum as a discrete sum, though for the free states, the tartarus AA code generates a continuum which must be integrated. Pauli blocking is accounted for explicitly in this method with the Fermi-Dirac thermal occupation factors, $f_{\alpha}$, in the numerator\footnote{Note this really comes from the retarded combination of polarizations combined with Fermi factors, $f_\beta(1-f_\alpha) - f_\alpha(1-f_\beta) = f_\beta - f_\alpha$.}. The last factor is the matrix element,
\begin{align}
    \<\phi_\beta| e^{-i \k \r} | \phi_\alpha \>  =  \int d^3\r e^{-i \k \r} \phi_\beta(\r)^* \phi_\alpha(\r).
\end{align}
Our states are single-electron orbitals from a spherically symmetric average atom, with normalizations given in
\cite{Blenski_1995},
\begin{align}
    \phi_{\alpha m l}(\r) &= R_{\alpha l}(r)Y^m_l(\hat{\r})\\
    \phi_{\alpha' m' l'}(\r) &= R_{\alpha' l'}(r)Y^{m'}_{l'}(\hat{\r})
\end{align}
Where $Y^m_l$ is a spherical harmonic with orbital angular momentum quantum number $l$ and the magnetic quantum number $m$, and $R_{\alpha l}$ encodes all radial information, where for bound states we will have $\alpha (\alpha') = n (n')$, and for the continuum $\alpha (\alpha') = E (E')$. We then expand the plane wave in terms of spherical harmonics and using a Clebsch-Gordan orthogonality relation and Uns$\ddot{o}$ld's theorem \cite{Arfken2013-wv}, we get the simplified answer for discrete bound-bound states
\begin{align}
    \chi^0_{bb}(k,\omega) &= g_s n_I \sum_{l,l',n,n'} \frac{f_{n'l'} - f_{nl} }{w - E_{nl} + E_{n'l'} + i \Gamma} \times \nonumber \\
    &  \sum_{\bar{l}} (2l+1) (2l'+1) C(l l' \bar{l} | 0 0 0 )^2 \times \nonumber\\
    & \left(\int dr j_{\bar{l}}(kr) R_{nl}(r) R_{n'l'} (r)\right)^2 \label{eq:chi_bb}
\end{align}
and for bound-free and free-bound contribution, we replace the discrete sum with an integral
\begin{align}
        \chi^0_{b\to f}(k, \omega) &= g_s n_I \sum_{l,l',n} \int dE' \frac{f_{nl} - f_{E'}}{w - E' + E_{nl} + i \Gamma} \times \nonumber \\
        & \sum_{\bar{l}} (2l+1) (2l'+1) C(l l' \bar{l} | 0 0 0 )^2 \nonumber \\ 
        & \left(\int dr j_{\bar l}(kr) R_{nl}(r) R_{E',l'} (r)\right)^2 \label{eq:chi_bf}
\end{align}
with the $f\to b$ contribution is gained by flipping the bound and free subscripts. 

We compute these matrix elements assuming a width sufficient to smooth out the AA continuum states. Since these states are defined uniformly in momentum space, the necessary numerical width in energy space is,
\begin{align}
    \Gamma(\omega) = c_\Gamma 2 \sqrt{\omega} \Delta \sqrt{E} 
\end{align}
 for $E$ a continuum energy and $\Delta \sqrt{E}$ a constant across continuum space, and $c_\Gamma$ a variable coefficient we set to one. Varying this parameter changes the expense of the code through the number of free states required to explicitly resolve, but otherwise does not change the result. Such a small but finite width vastly simplifies the numerical integration of the dielectric around the plasmon pole, avoiding a tedious splitting between the near-singular plasmon and the continuum though this is possible\cite{hentschel2023improving}. 
  
Treating the $f-f$ contribution in this way is needed for a fully consistent theory, yet doing this comes with conceptual problems\cite{CAIZERGUES20167, johnson2009low}, and requires an expense of $\mathcal{O}(N_E^2 N_l^2 N_r)$ compared to $\mathcal{O}(N_E N_b N_l N_r)$ for the bound-free case. Since we typically require thousands of free points and an $N_l\sim 10-100$, this method is intractable without a different approach. We thus assume the finite-temperature Lindhard response for the free-free computation. One could instead attempt to improve this result using the AA density of states, as in \cite{hentschel2023improving}, but we find this is problematic when there are strong resonances in the continuum, see Appendix~\ref{app:Lindhard_improved_DOS}.


\begin{figure*}[t!]
    \centering
    \begin{minipage}{0.5\textwidth}
        \centering
        \includegraphics[width=1.0\textwidth]{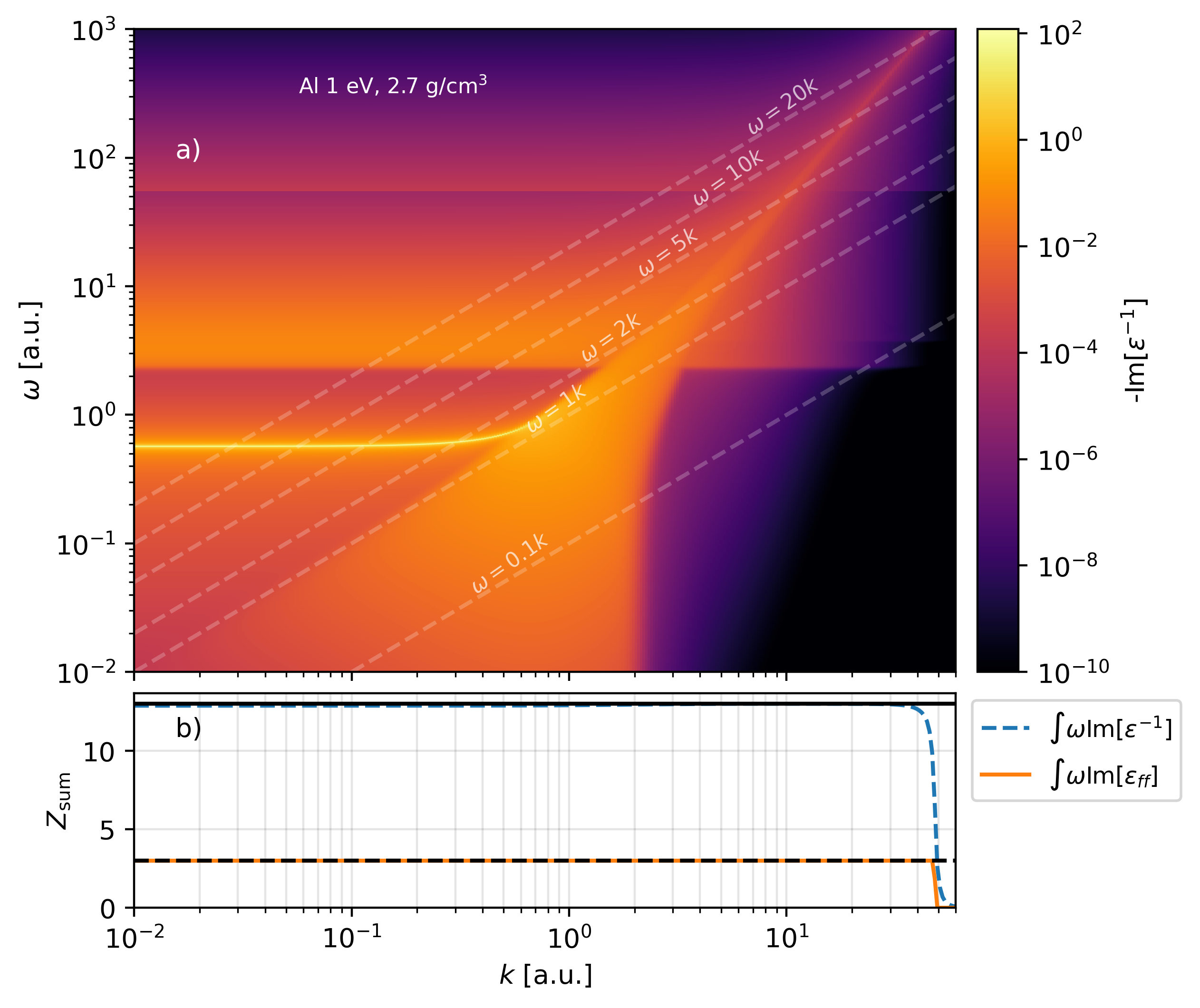} \\ 
    \end{minipage}%
    \begin{minipage}{0.5\textwidth}
        \centering
        \includegraphics[width=1.0\textwidth]{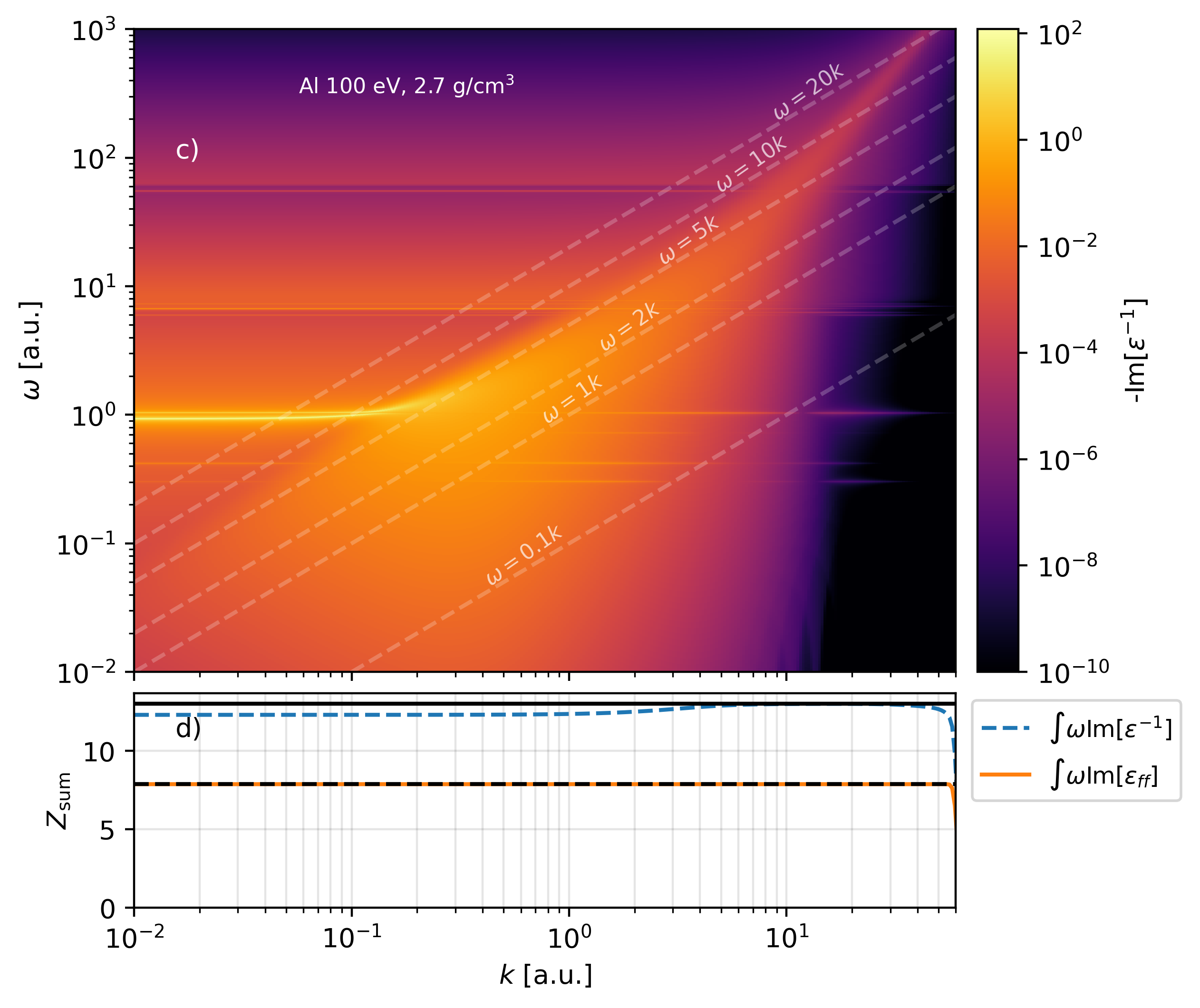} \\ 
    \end{minipage}
    \caption{In {\bf a)} we show the ELF for aluminum as a function of the momentum and energy transferred at $T=1$ eV and $\rho=2.7$ g/cm$^3$, with $\Zbar=3$. In white dashed lines we show the integration limits, $\omega \leq v_i k$, for several ion velocities relevant to stopping see Eq.~\eqref{eq:SP}. In {\bf b)} we show the corresponding f-sum rule\cite{Kremp2004-kr} as a function of $k$ for the full dielectric (blue) compared to $Z$ (black solid) the and f-f part only (orange) versus $\Zbar$. In {\bf c)} we have the ELF now for $T=100$ eV, $\rho=2.7$ g/cm$^3$ aluminum with $\Zbar=7.88$. In {\bf d)} are the f-sum rules, showing a consistent deficit in the all-electron one at lower $k$. The large drop off in sum rule at very high $k$ is purely a grid limitation to finite $\omega$ which stopping is converged against. }
    \label{fig:Al_ELF_grid}
\end{figure*}

\subsection{Resulting energy loss functions}
The relevant quantity for stopping power is the energy loss function (ELF), given in Eq.~\eqref{eq:ELF}. This is obtained using the polarization functions, Eqs.~\eqref{eq:chi_bb},\eqref{eq:chi_bf} with tartarus code bound and continuum orbitals, and the finite-temperature Lindhard function for the free-free contribution. 

The ELF for Aluminum at solid density is shown in Fig.~\ref{fig:Al_ELF_grid}.  On the left in panel {\bf a)} we have the $T=$ 1 eV case where we can clearly see the sharp plasmon peak at $\omega = 0.58$ a.u. with an upturn just before reaching the continuum of individual Fermi-sphere electron scattering losses\cite{Kremp2004-kr}. We can see two distinct bound-free ionization thresholds which are roughly $k$ independent except near the narrow impact approximation asymptote at $\omega=k^2/2$ which appears as a sharp feature extending to the upper right. In panel {\bf c)} of Fig.~\ref{fig:Al_ELF_grid} we have Aluminum at $T=$ 100 eV. We see largely the same features but with a large smearing from the higher temperature and additionally the appearance of new bound-bound lines. These bound-bound lines extend into the high $k$ regime only decaying around when $k\gtrsim 1/r_{b_i}$ for $r_{b_i}$ the size of the $i$-th bound state. Since these lines extend to very high momentum, they fall under the low velocity $\omega= v_i k$ lines and will later contribute to low velocity stopping power.

To check the validity of our cmRPA method and verify convergence, we check the following f-sum rules on the dielectric \cite{Mahan2010-os},
\begin{align}
    \int^\infty_0 d \omega \omega \Im[\epsilon(k,\omega)] &= 2\pi^2 n_e\label{eq:sum_diel_1}, \\
    \int^\infty_0 d \omega \omega \Im[\epsilon^{-1}(k,\omega)] &= -2\pi^2 n_e \label{eq:sum_diel_2}.
\end{align}
In the lower part of Fig.~\ref{fig:Al_ELF_grid}, in panels {\bf b)} and {\bf d)}, we see the $k$ dependence of the f-sum rules for the full ELF (blue dashed) which is normalized so that it should sum to $Z=13$ (black solid). We see a small violation of this sum rule on the order of $1\%$. The reason is the inconsistency between our Lindhard assumption for free-free and the AA-derived response for bound-bound and bound-free. In particular, Lindhard assumes plane wave eigenstates as free the free electrons, which is inconsistent with the AA Hamiltonian that defines the orbitals used in the b-b and b-f calculation. Another way to say this is that in general the sum rule contribution is only defined over all electrons, and thus spectral weight can move between free and bound contributions, but here we use Lindhard free-free(orange solid) which can only ever sum to the input free electron density, here $\Zbar=3$ (panel \textbf{b)}) or $\Zbar=7.88$ (panel \textbf{d)}) shown as black dashed lines. The Lindhard f-sum rule to $\Zbar$ is satisfied within $0.1\%$ in all plots. 

Despite the small violation in all electron f-sum rule, at high $k$ we see convergence to the full sum rule due to the Lindhard plane waves becoming increasingly similar to the high energy continuum scattering states in the AA at higher $k$ and $\omega$. Thus, the violation is numerically largest in the long wavelength limit that has little impact on stopping power. In Fig.~\ref{fig:Al_ELF_grid} the all-electron f-sum rule is violated more in the $T=100$ eV case of panel \textbf{d)} due to a sharp resonance in the density of states. This sharp resonance corresponds to a quasibound 3d state that is strongly different from the plane-waves Lindhard implicitly assumes. 

\begin{figure*}[t!]
    \centering
    \begin{minipage}{0.48\textwidth}
        \includegraphics[width=1\textwidth]{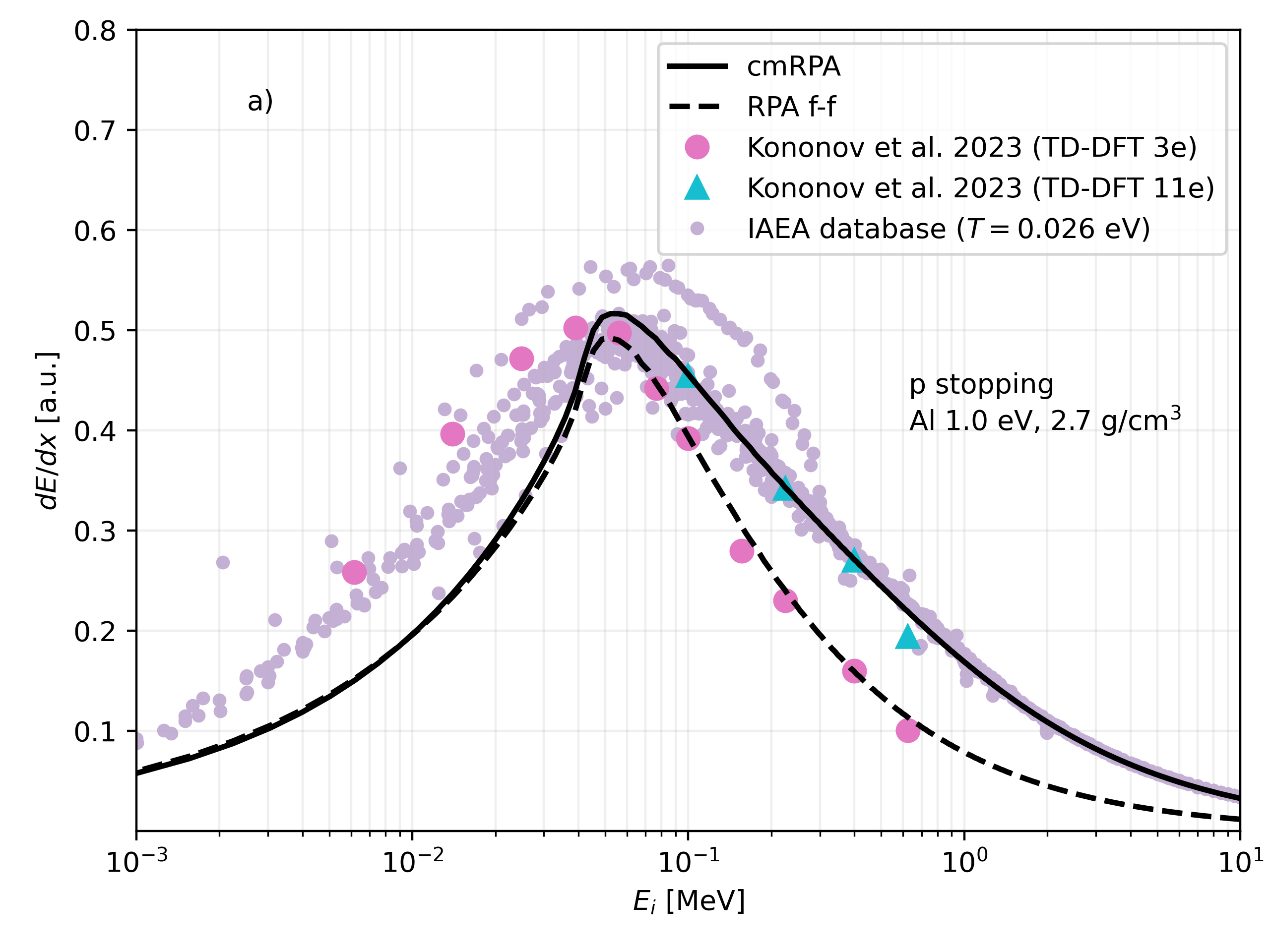}
        \includegraphics[width=1\textwidth]{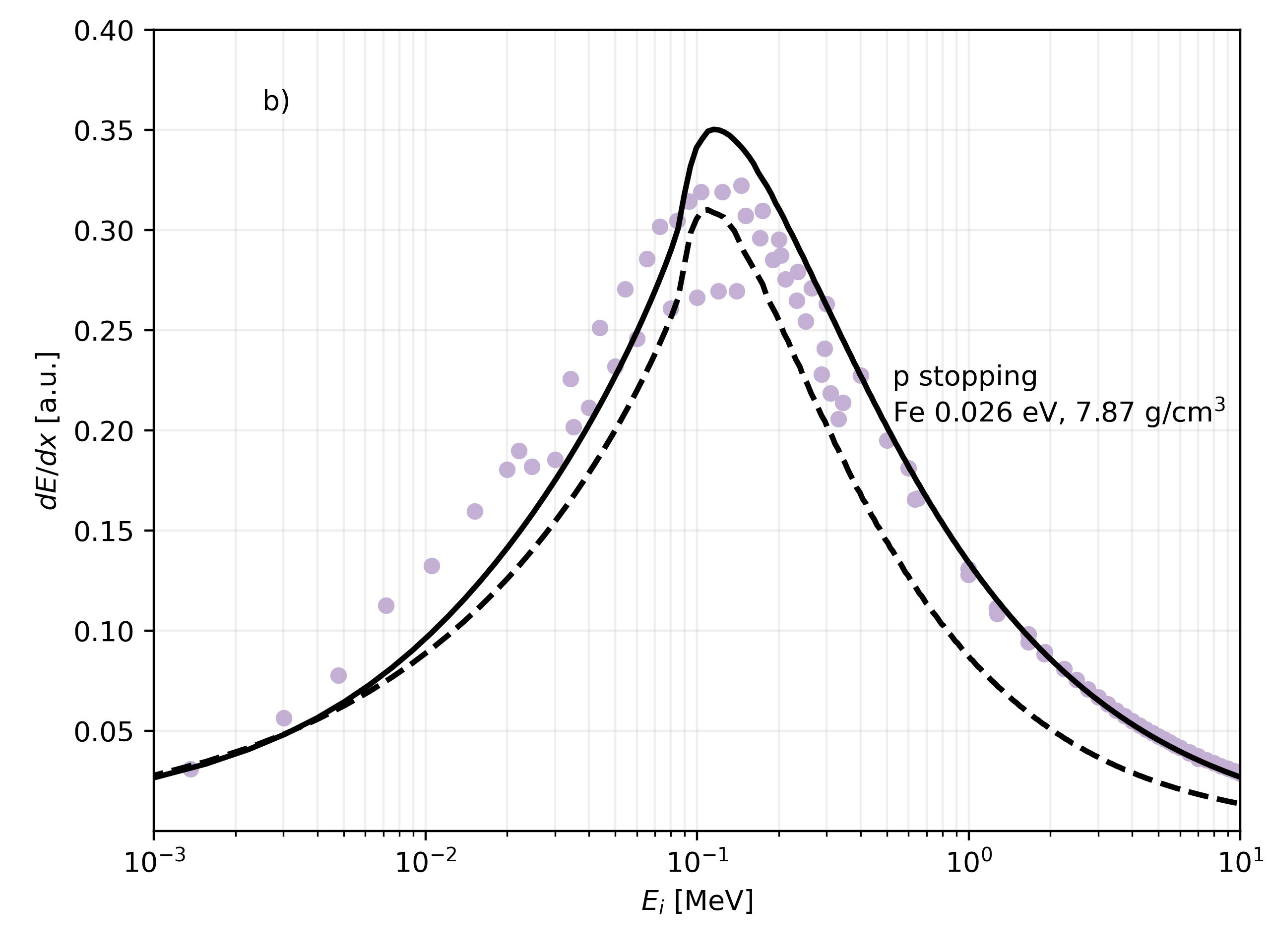}
    \end{minipage}%
    \begin{minipage}{0.48\textwidth}
        \centering
        \includegraphics[width=1.0\textwidth]{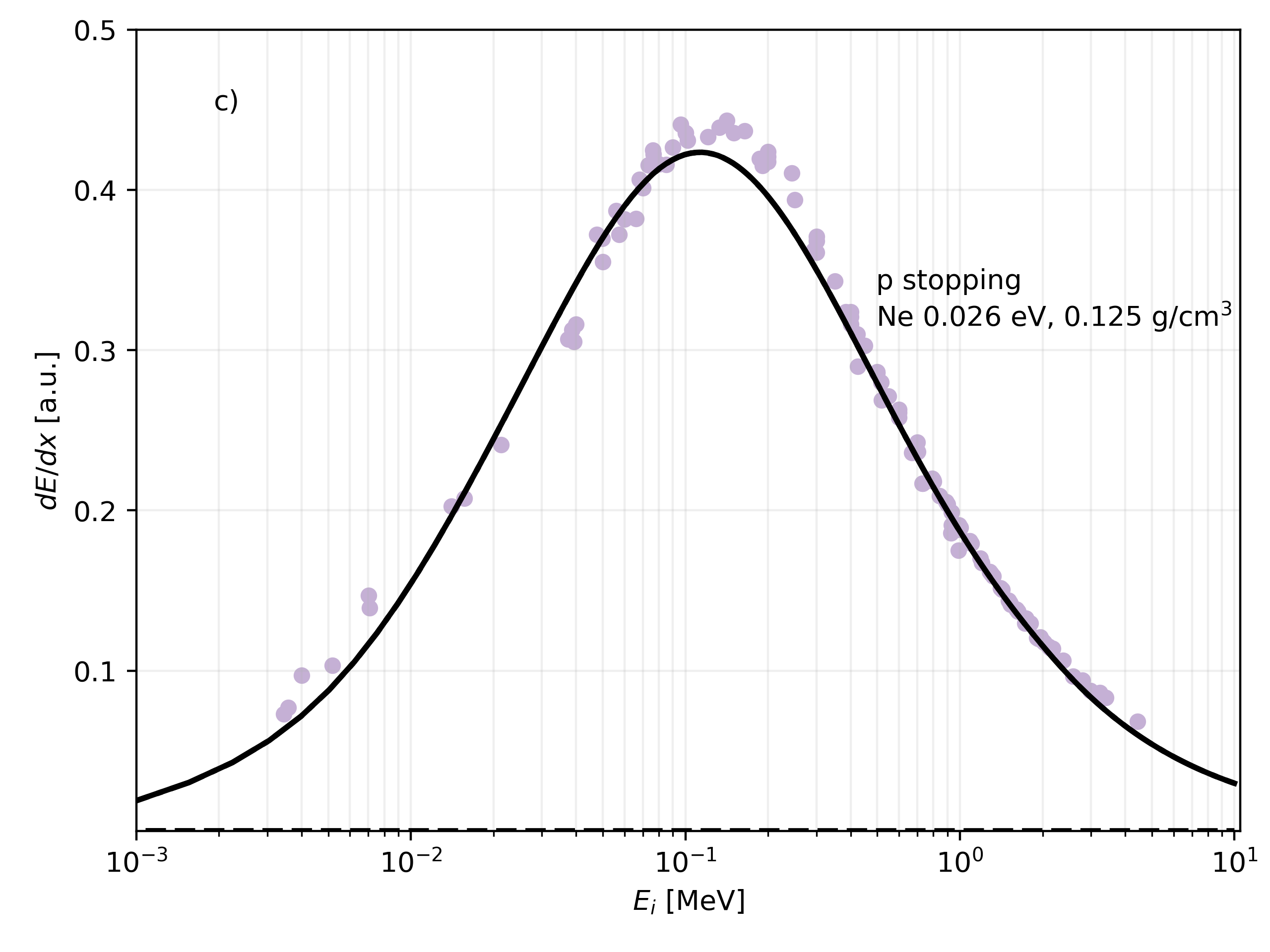}
        \includegraphics[width=1.0\textwidth]{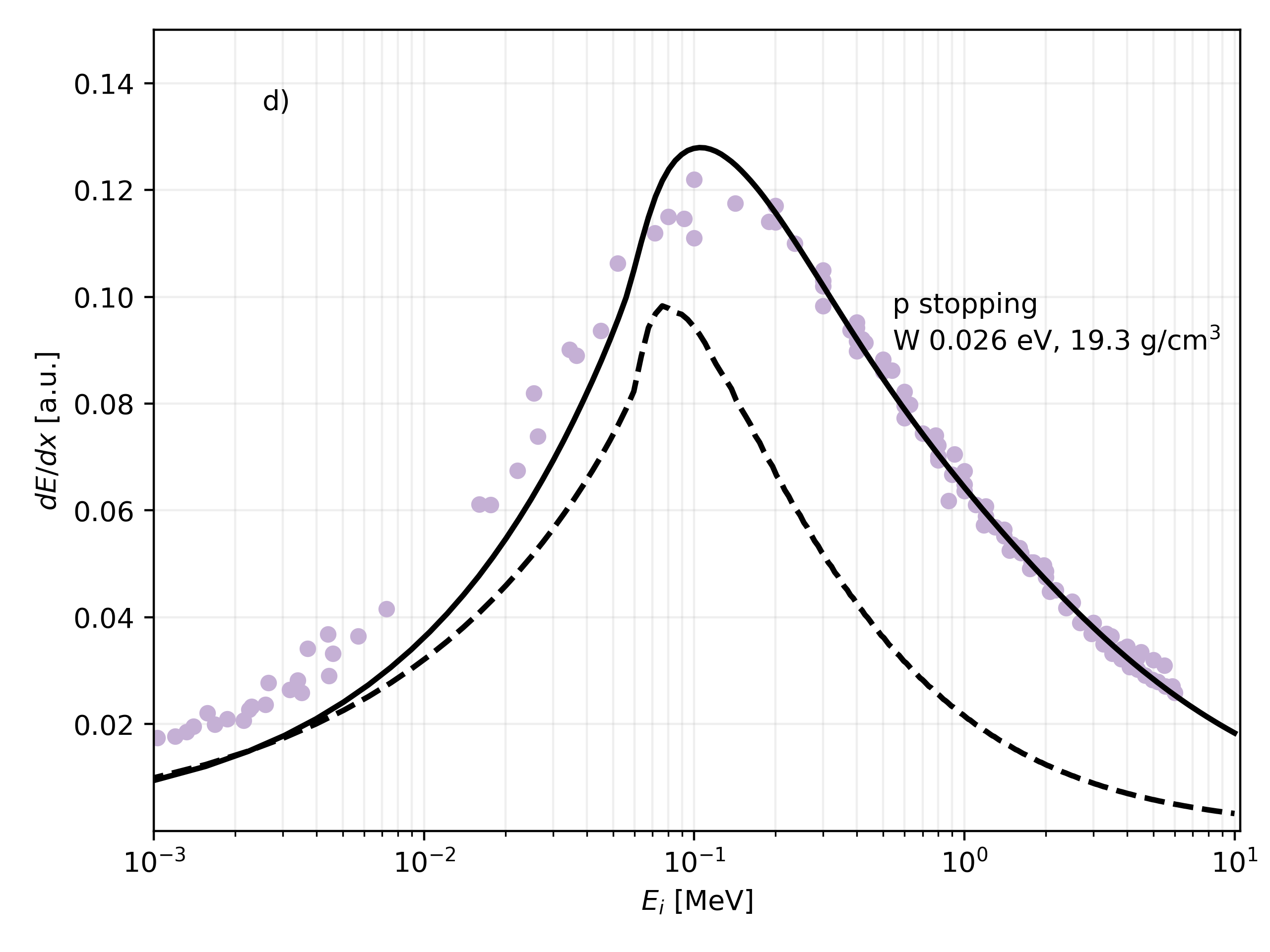}
    \end{minipage}
    \caption{  We plot proton stopping near ambient conditions from the IAEA experimental database (purple circles) \cite{IAEA}, the f-f contribution from RPA with $\Zbar=3$ (black dashed), the b-f (black dash-dot) in target materials and the model presented in this paper, cmRPA (black solid). We have {\bf a)} aluminum at ambient $\rho=2.7$ g/cm$^3$ and $T=1$ eV where we also have TD-DFT data \cite{Kononov2023-wl} for a 3e pseudopotential (pink circles) and an 11e pseudopotential (turquoise triangles), {\bf b)} iron at ambient density, $\rho=7.87$ g/cm$^3$ and $T=0.026$ eV, {\bf c)} neon gas at $\rho=0.125$ g/cm$^3$ and $T=0.026$ eV,  {\bf d)} tungsten at ambient conditions, $T=0.026$ eV and $\rho=19.3$ g/cm$^3$. 
    }
    \label{fig:ambient_benchmark}
\end{figure*}

Lastly we note that this discussion on sum rules applies when the radial integrals appearing in Eqs.(\ref{eq:chi_bb}, \ref{eq:chi_bf}) are taken over the entire extent of the bound state, which can be several times the ion-sphere radius. As an alternative model, we can integrate only out to the ion-sphere radius. This is computationally expedient but results in additional sum rule violations since the states are mutually orthogonal only when integrated over all space. However we find the impact on stopping is negligible, so in the rest of the paper we use truncated integrals, see Appendix~\ref{app:truncated}.


\section{Stopping Power Results} \label{sec:results}
In this section we show results from our orbital based linear response stopping power approach. Since our approach is in the linear response framework, we expect our model to perform well only above the Bragg peak where stopping power is at a maximum. First we validate our model by comparing to the wealth of ambient experimental data and available TD-DFT simulations, showing very good agreement. We then continue with application to the WDM and high energy density (HED) plasma regimes.

\subsection{Comparison with Experiment and Simulation in Ambient Conditions}
 We validate our stopping model in Fig.~\ref{fig:ambient_benchmark} for four different materials at near ambient conditions, comparing to experimental data compiled in the IAEA database\cite{IAEA}. In all panels we show the $T=$ 0.026 eV experimental data (purple dots), our full model accounting for bound and free states in the restricted ion sphere (solid black), and pure RPA assuming $\Zbar$ electrons (dashed black). In this paper, RPA will always refer to assuming the number of electrons given by $\Zbar$ from our AA model, which is equivalent to the nuclear charge minus the number of occupied bound states. For comparisons to other stopping models in the literature, see Appendix ~\ref{app:extra_ambient}.

In all four cases of Fig.~\ref{fig:ambient_benchmark} we show very good agreement of our stopping model with experimental data above the Bragg peak. At the peak, we slightly over-predict the stopping power for the cases of aluminum, iron and tungsten due to the sharpness of the plasmon peak in the free-electron Lindhard response. Below the Bragg peak, our linear response limit does not capture stronger binary scattering as would be better captured in T-matrix based models such as \cite{gericke1996stopping} or more recent improvements using the potential of mean force \cite{Babati_2026}, and so typically under-predicts the stopping there, see \cite{kononov2025nonlineareffectslightionstopping}. There are also electron gas correlation effects missing, which can be captured through local field corrections\cite{Faussurier2025}. We show one example of this correction in Appendix ~\ref{app:LFC}.

The accuracy of our approach to bound state stopping is demonstrated in the upper right case of Fig.~\ref{fig:ambient_benchmark} where we show a neutral Neon gas at room temperature that has no free-free component. In this case we see even better agreement with the data than in other cases, supporting the accuracy of our bound state model. It is worth noting that despite the general expected decrease in accuracy of linear response below the Bragg peak, here where only bound-free response is present, we accurately predict experiment for both low and high velocities. Despite this agreement, we note that in lower $Z$ gases, our underlying AA model becomes inaccurate due to self-interaction errors in DFT, and for those cases Hartree-Fock based orbitals might give a better stopping power. 

In panels {\bf a)}, {\bf b)}, {\bf d)} of Fig,~\ref{fig:ambient_benchmark}, we note there is a sharp feature in the bound-free contribution that can be seen most clearly around the energy of the Bragg peak in the case of ambient tungsten. This sharp change is a consequence of the mixing between bound-free and free-free transitions, see Section~\ref{sec:channel_mixing_effect}, and occurs exactly where the undamped plasmon response turns on. In the Lindhard approximation for the free-free response the plasmon peak is very sharp. If the plasmon peak is broadened via higher temperatures or scattering, then this feature would be smoothed out.

\subsection{Temperature Dependence of the full Stopping Power}
The back-of-the-envelope free electron binary scattering calculation for stopping power gives two basic regimes, with a peak at the average or thermal electron velocity,
\begin{align}
    \frac{dE}{dx} \sim
        \begin{cases}
        Z_b^2 n_e v_i/v_e^3 & \text{if } v_i \ll v_e \\
        Z_b^2 n_e/v_i^2 & \text{if } v_i \gg v_e.
        \end{cases}
\end{align}
for free electron density $n_e$ and averaged thermal velocity $v_e$ which at low temperatures becomes the Fermi velocity.

\begin{figure}[t!]
    \centering
    \includegraphics[width=1\linewidth]{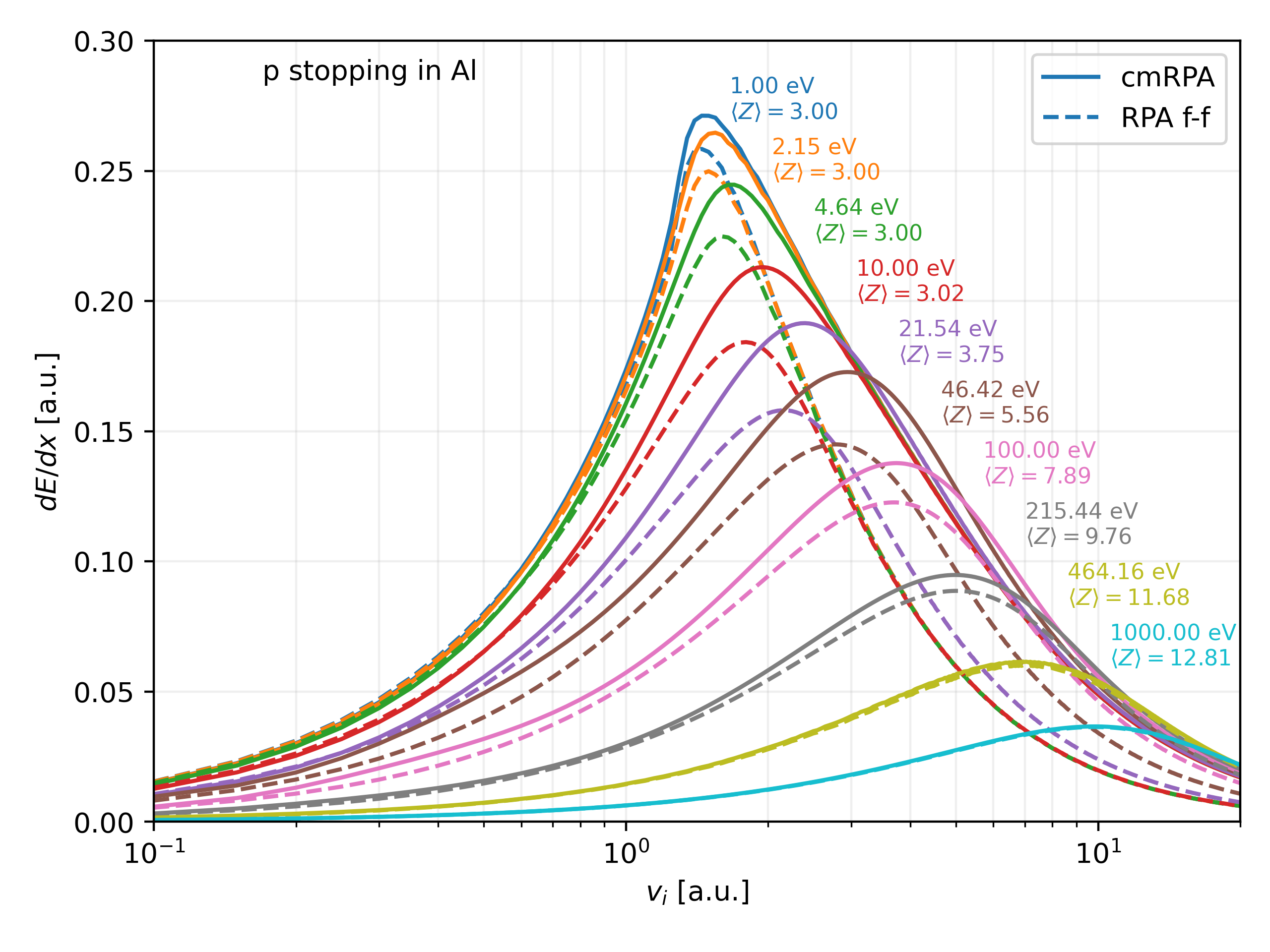}
    \caption{Stopping power of our full model (solid) and RPA only (dashed) for solid density aluminum at temperatures from 1 eV to 1 keV.}
    \label{fig:Al_T_scan}
\end{figure}

This is a crude approximation, but it captures the correct asymptotic scaling for both dielectric and binary scattering theory up to the Coulomb log and assuming only free-free scattering. We can see some of this behaviour in Fig.~\ref{fig:Al_T_scan} with our full model (solid) compared to free-free only (dashed). We show solid density aluminum at varying temperatures from essentially zero temperature to 1 keV. The lower velocity side of the stopping power and the height of the Bragg peak is highly sensitive to temperature whereas the high velocity limit is largely only dependent on the charge state. The increase of the width of the Bragg peak at higher temperatures corresponds to a widening of the plasmon peak as the high temperature smears out the Fermi occupation of states. 

While the Bragg peak shifts to higher energies as temperature is increased, the binding energies are dependent on this temperature change only weakly through the temperature dependence of the free electron screening length. Thus, the ion energies relevant for the b-f contribution remain nearly constant as temperature increases. For aluminum this means at cold temperatures the primary impact of bound-free is above the Bragg peak, but by a temperature of $10-30$ eV, the Bragg peak occurs at the same energies as the bound-free excitation, resulting in modifications to stopping primarily at the Bragg peak. This is seen as a larger change in peak height for those conditions in Fig.~\ref{fig:Al_T_scan}. As temperature is further increased, ionization increases, bound states are ionized, and the pure free-free (dashed) lines converge to the full cmRPA lines (solid).  


\subsection{The Effect of Channel Mixing and Bound Stopping}\label{sec:channel_mixing_effect} 
\begin{figure}
    \centering
    \includegraphics[width=1\linewidth]{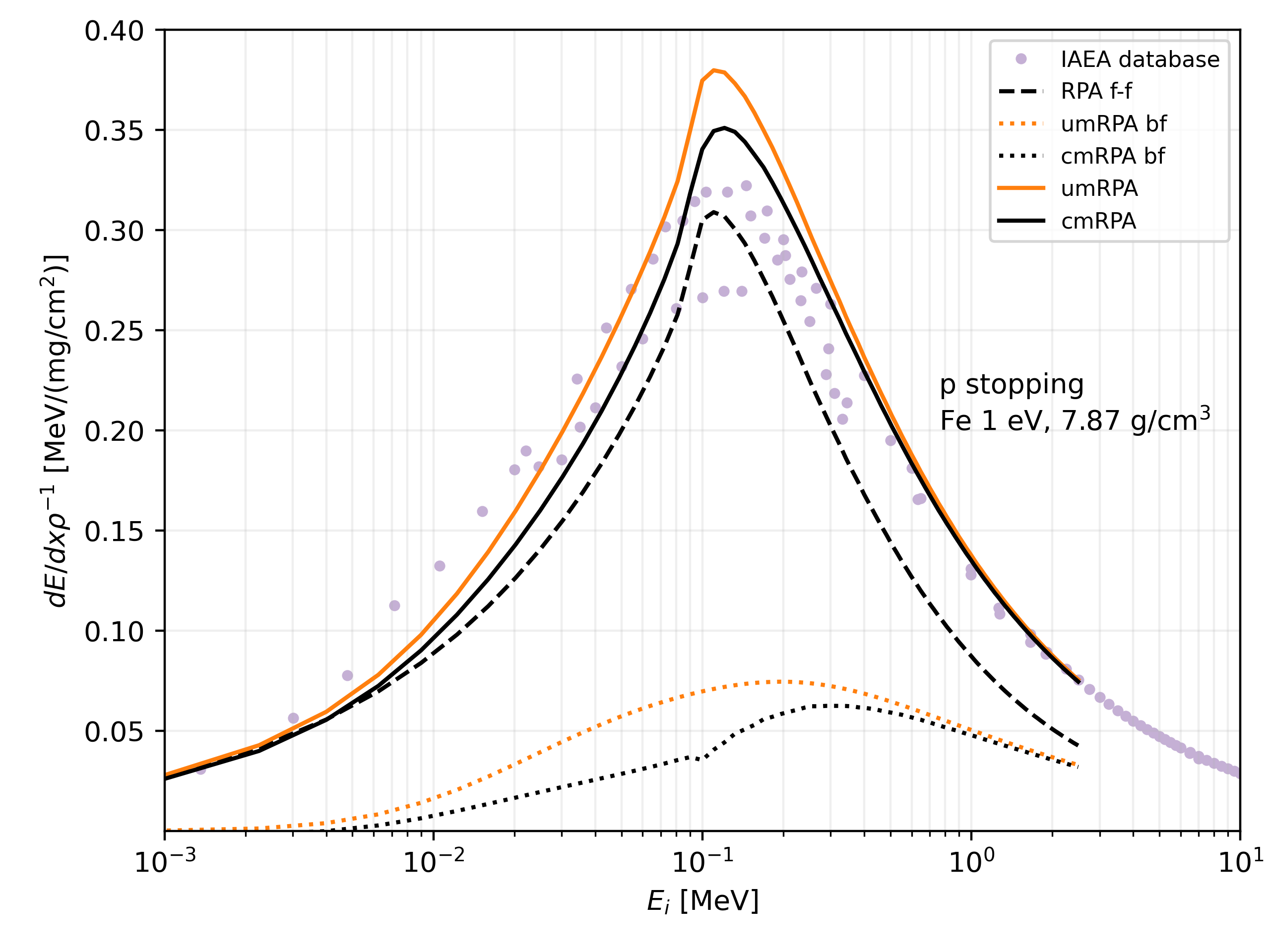}
    \caption{We show the effect of hybridization of the plasmon response with the bound-free response on the stopping in iron at a temperature of 1 eV and solid density, 7.87 g/cm$^3$. We show our stopping model prediction in the mixed cmRPA ELF model (black solid) corresponding to Eq.~\eqref{eq:ELF} with bound-free contribution (black dotted) versus the unmixed version using Eq.~\eqref{eq:ELF_unmixed} (orange solid) and unmixed bound-free contribution (orange dotted), and the RPA or f-f contribution (black dashed). }
    \label{fig:Fe_mixing}
\end{figure}

The natural form of the dielectric where all bound and free transitions are mixed together in the cmRPA approximation, defined by Eq.~\eqref{eq:ELF}, causes the channels to interfere with each-other\footnote{This interference is not quantum interference at the level of wavefunctions which would exist in a more general model, but classical interference of the resulting individual response terms.}. This means the bound contribution to stopping will in general depend on the location and shape of the plasmon peak, as well as the screening by the free continuum. We thus define the bound stopping contribution from a channel by the difference between the stopping with and without that state,
\begin{align}
    \frac{dE}{dx}{\Big |}_{bf}^{\rm cmRPA} &= \frac{dE}{dx}\left(\chi^0_{ff} + \chi^0_{bf} \right) - \frac{dE}{dx}\left(\chi^0_{ff}\right), \label{eq:SP_bf_definition}
\end{align}
which we clarify will not generally be the same as the stopping power computed with the response from that state only, as in Eq.~\eqref{eq:ELF_unmixed}. We show the impact of this difference in Fig.~\ref{fig:Fe_mixing}. The cmRPA method causes a response interference between the plasmon peak at 1.1 Ha and the 3p binding energy of 1.5 Ha and thus a decrease in the stopping (black) relative to the unmixed version (orange) for both the total stopping (solid) and bound contribution alone (dotted). In cases where the various response channels are well separated, the difference between these methods goes to zero. We also see that at the large energies where higher frequency response becomes dominant, the mixed and unmixed responses become identical. This is because frequencies above the threshold to excite both plasmon and bound-free responses constructively interfere rather than destructively.   
We can extend the idea of Eq.~\eqref{eq:SP_bf_definition} to find the contribution of individual bound state $i$,
\begin{align}
    \frac{dE}{dx}{\Big |}_{b_i}^{\rm cmRPA} &= \frac{dE}{dx}\left(\chi^0_{ff} +  \chi^0_{b}  \right) - \frac{dE}{dx}\left(\chi^0_{ff} + \chi^0_{b\neq i}\right),\label{eq:SP_bound_i}
\end{align}
where $\chi^0_{b} = \chi^0_{bf} + \chi^0_{bb}$ is the total bound contribution, and $\chi^0_{b\neq i}$ is equivalent except bound state $i$ is not included. This equation also coincides with the difference between a TD-DFT stopping power computation with bound states frozen, and an all electron calculation \cite{osti_2431819}. 

\begin{figure}
    \centering
    \includegraphics[width=1\linewidth]{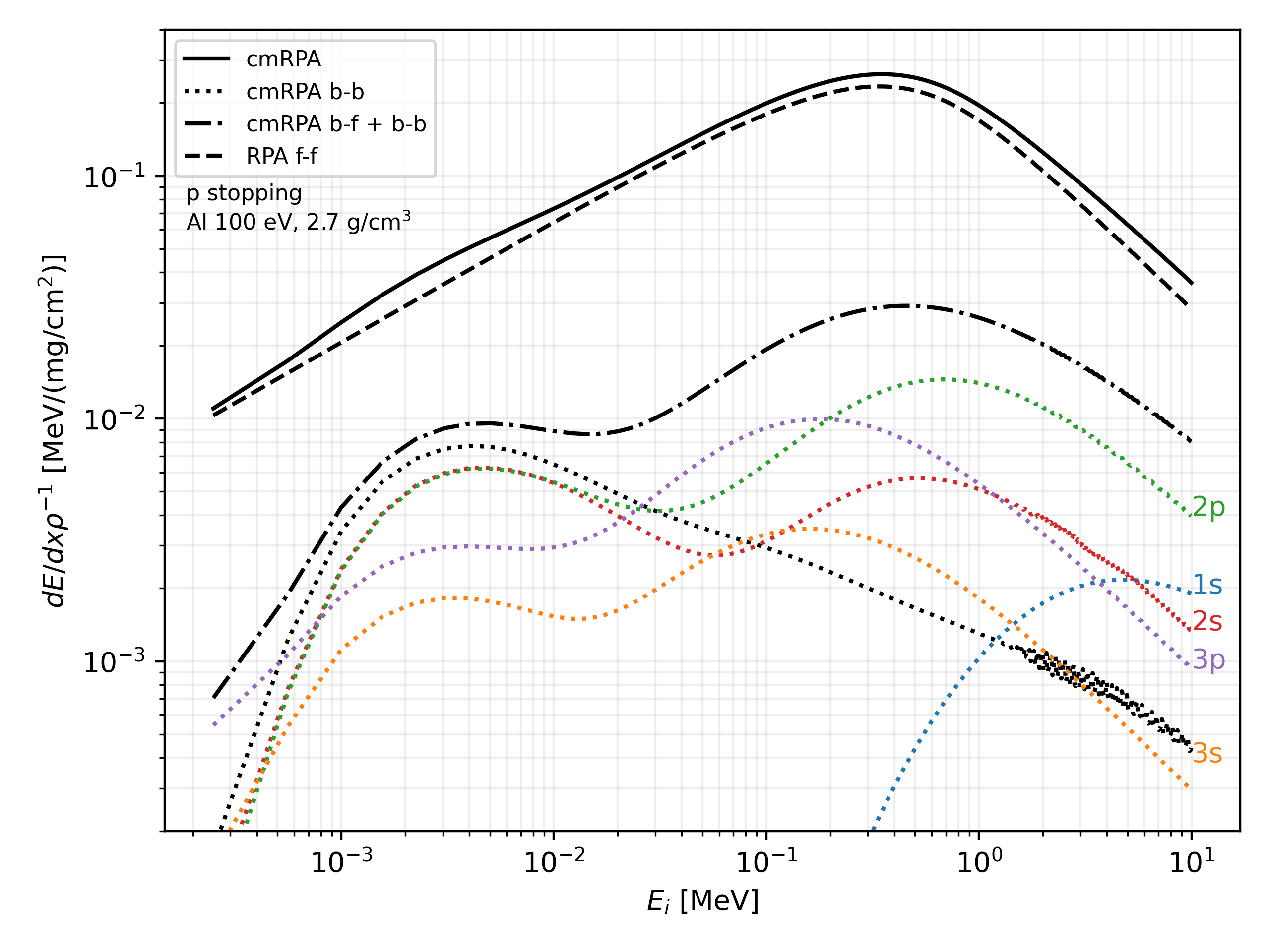}
    \caption{Proton stopping in aluminum at a temperature of 100 eV and solid density, 2.7 g/cm$^3$. We show our stopping model prediction (black solid), the RPA or f-f contribution (black dashed), as well as the total b-b (black dotted), total b-f + b-b (black dash-dot) and the contribution from individual states including b-b and b-f labeled and colored as in the inset. }
    \label{fig:Al_100eV_contributions}
\end{figure}

In Fig.~\ref{fig:Al_100eV_contributions} we show the stopping power of Al at $T=100$ eV by the individual contribution of each state using Eqs.~\eqref{eq:SP_bound_i}, \eqref{eq:SP_bf_definition} including bound-free and bound-bound response, the latter of which only show up at higher temperatures or lower densities. We show the full stopping (black solid), the full free-free contribution (black dashed), the full bound contribution (black dash-dot) and the bound-bound contribution (black dotted). Note for free-free we do not use the subtraction procedure in Eq.~\eqref{eq:SP_bf_definition}, making this just the Lindhard RPA response. Each bound state has its own contribution that includes all bound-bound and bound-free channels it is involved in (colored lines). 
We see a bound-bound contribution at low velocity due to the visible response at large $k$ and low $\omega$ in Fig.~\ref{fig:Al_ELF_grid}, and we see the bound-free contribution generally at high energy starting at the lowest ionization energy. Thus in terms of the per-state contribution, we see two peaks in all cases except for 1s where bound-bound and bound-free transitions occur only at very high energies.  


\subsection{High Energy Density Tungsten }
In this section, we consider the stopping power for highly compressed and high temperature cases relevant for inertial confinement fusion applications, showing that our model gives reasonable results even in extreme conditions. 

A particular example is for double-shell configurations where a high-Z element such as tungsten is used as a high inertia pusher onto the DT gas below \cite{10.1063/1.5042478}. During peak compression, this can reach densities on the order of 1000's of grams per cubic centimeters and multi keV temperatures. At these conditions the atoms of the DT gas and most materials would be completely ionized, and thus our stopping model would reduce exactly to RPA stopping. However, the high-Z pusher will not be fully stripped unless the temperature is a few 10's of keV due to the very deeply bound core states \cite{NIST_ASD}. The exact temperature will strongly depend on the extent of fuel-pusher mixing and the location inside the tungsten layer, ranging from 100's of eV to 10's of keV \cite{hayes2025reaction}.  

\begin{figure*}[t!]
    \centering
    \begin{minipage}{0.48\textwidth}
        \includegraphics[width=1\textwidth]{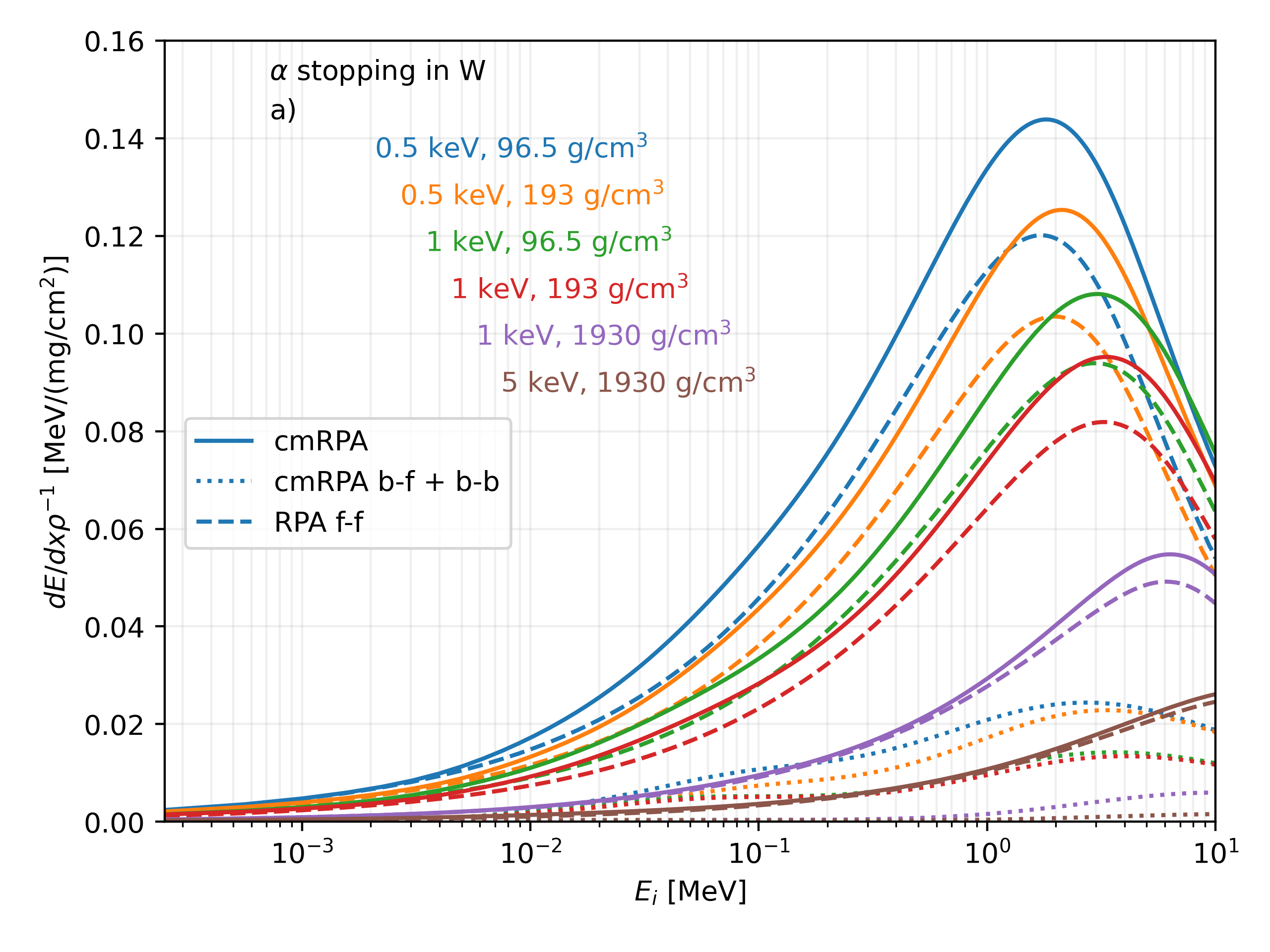}
    \end{minipage}%
    \begin{minipage}{0.48\textwidth}
        \centering
        \includegraphics[width=1.0\textwidth]{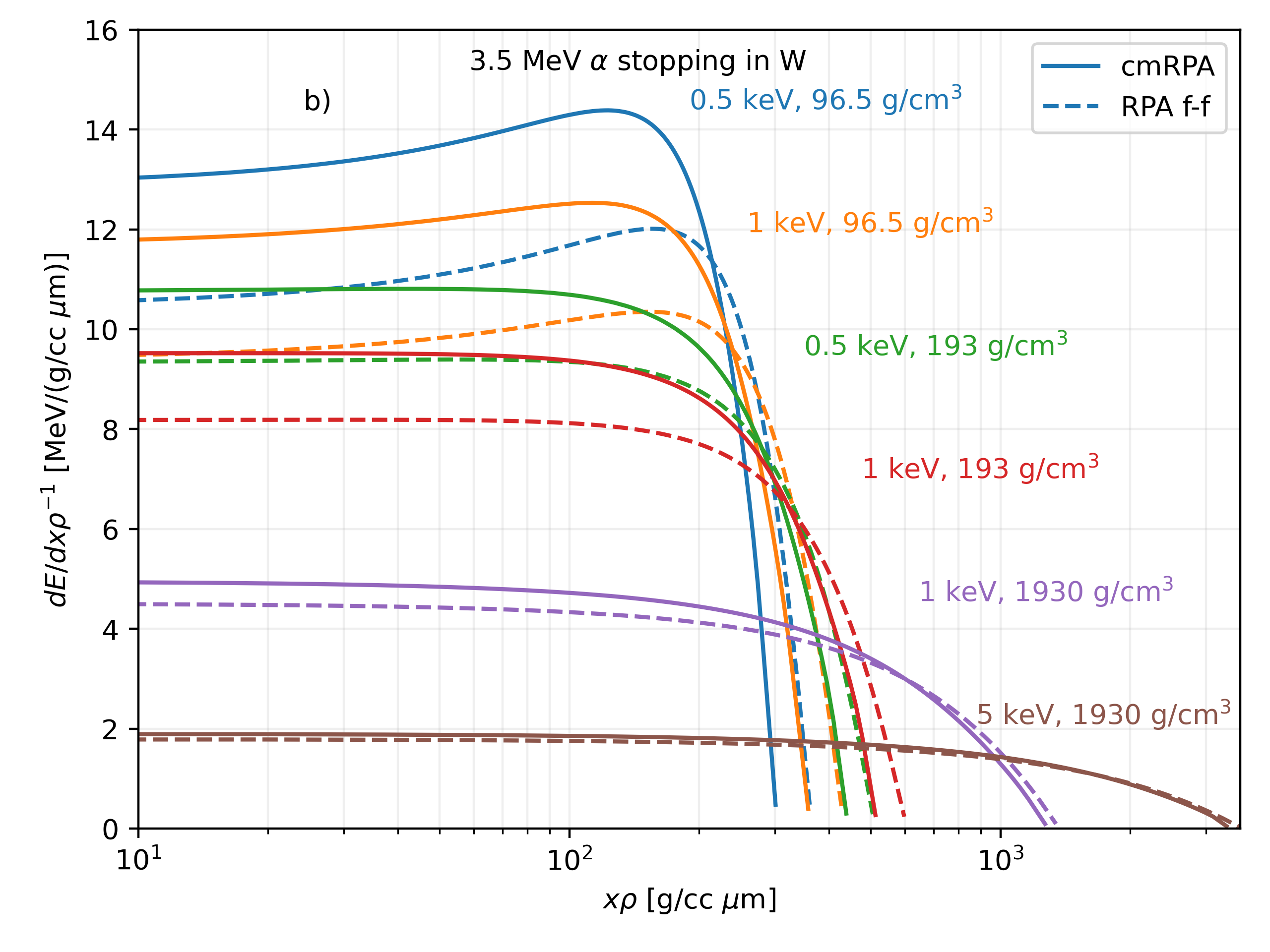} \\ 
    \end{minipage}
    \caption{ {\bf a)} Alpha stopping in tungsten at six ICF relevant conditions shown in the inset and {\bf b)} the corresponding energy deposition rate over areal density position for the same points. Typical areal densities in double-shell experiments are on the order of several thousand $\rm{ g/cm^3 }\ \mu \rm{m}$\cite{hayes2025reaction}. Here the calculated $\Zbar$ is, from top to bottom (blue to brown), 34.4, 43.8, 34.5, 43.1, 48.3, and 62.  }
    \label{fig:W_comparison}
\end{figure*}

The stopping power of tungsten in conditions relevant to these experiments are shown in Fig.~\ref{fig:W_comparison}. On the left we show stopping power for a variety of highly compressed, hot conditions potentially relevant to heating double-shell \cite{10.1063/1.5042478} capsules and stopping related diagnostics \cite{hayes2025reaction}. On the left of Fig.~\ref{fig:W_comparison} we see the stopping power of our full model (solid) versus the RPA contribution (dashed) normalized to partially remove the density dependence. We see the colder, less dense conditions have a significant bound contribution and a lower Bragg peak. As both density and temperature increase, the thermal electron velocity increases and the Bragg peak shifts to higher energies and broadens as the large temperature smears out response features. The magnitude, once normalized by density, also decreases. We can see from comparing the bound contribution from, for example blue to orange, that density has little effect on the bound contribution, but decreases the free-free contribution, from increased Pauli blocking and the faster plasmon response decreasing the maximum impact parameter. One can see the effect of temperature in the low velocity limit where the electrons are fast, in which case the cross section decreases with increasing temperature, lowering and shifting the Bragg peak.  

On the right panel of Fig.~\ref{fig:W_comparison}, we see the corresponding energy deposition of a 3.5 MeV DT fusion produced alpha particle along a trajectory parameterized by areal density. We can see that the highly compressed cases at 100 times compression (purple and brown), relevant to near peak burn time, have Bragg peaks above the alpha energy, giving a substantially longer range than the other cases. We note that for a typical tungsten layer in a double-shell capsule of $\sim 40 \mu {\rm m}$ wide, the areal density is initially $\sim 700\ \mu {\rm m g/cm^3}$. The areal density then increases as the capsule is compressed by a factor of the convergence ratio squared $(R/R_0)^2$ for initial and final tungsten radius $R_0$, $R$, respectively. Based on this, a DT generated alpha would likely pass through a small portion of the hot tungsten layer, but unless asymmetries create a low local areal density, the tungsten will always stop the $\alpha$ particles eventually. However the stopping distance changes drastically between the cold and hot tungsten implying that injections of cold tungsten may have an observably different impact on alpha stopping than mixed hot regions.       

\begin{figure}
    \centering
    \includegraphics[width=1\linewidth]{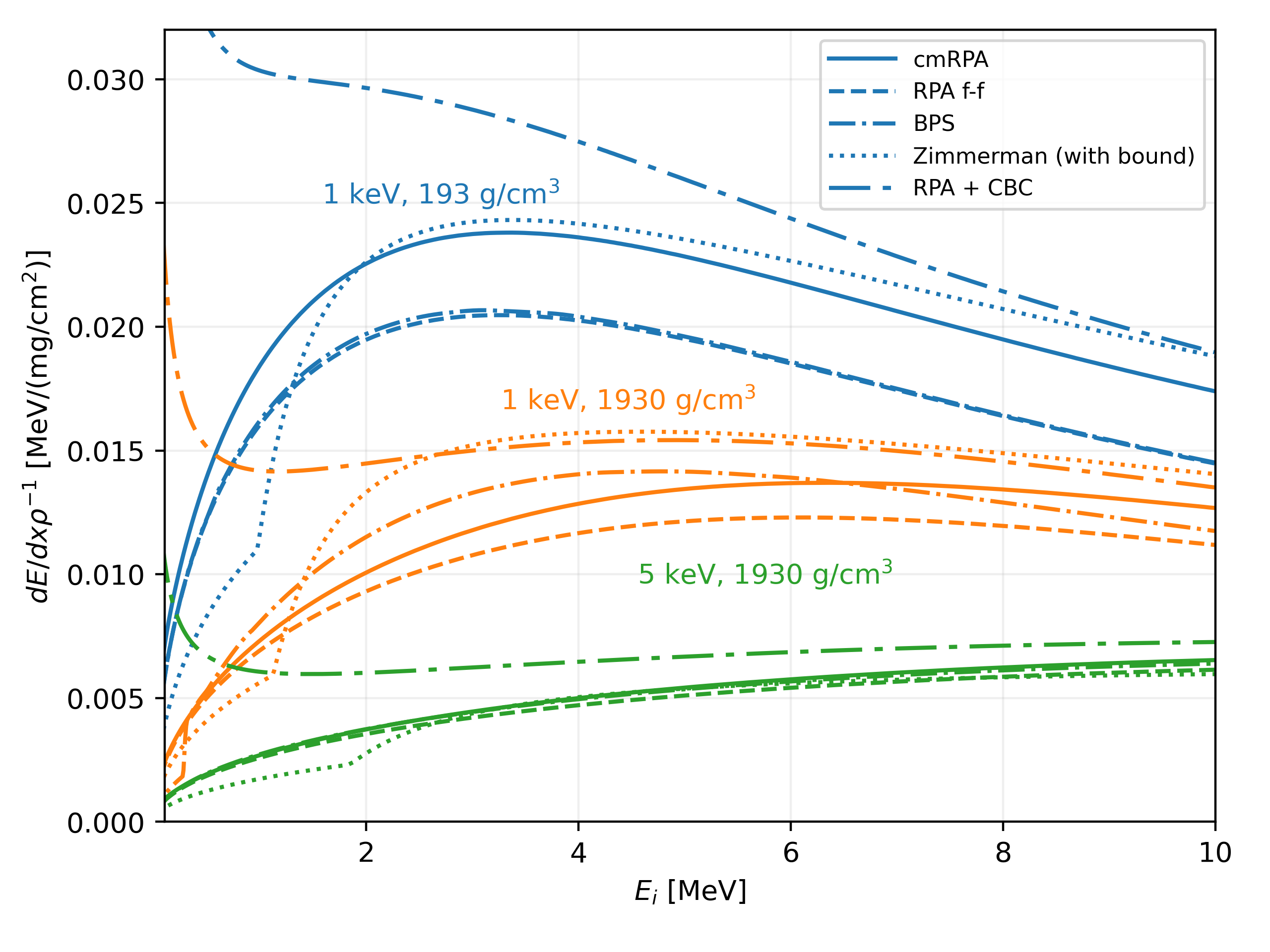}
    \caption{ For HED tungsten at 3 conditions in the range of 10-100 times compressed (conditions in the inset), we compare our model (solid) with RPA (dashed), BPS (dash dot), Zimmerman (dotted), and CBC (long dash dot). We can see large model variation in these conditions, but by a temperature of $5$ keV, the cmRPA model becomes almost identical to the pure free electron RPA due to the near complete ionization.}
    \label{fig:HED_model_comparison}
\end{figure}

We additionally do a model comparison between our results and other models, shown in Fig.~\ref{fig:HED_model_comparison}. In blue the tungsten is only 10x compressed, and the high keV temperature causes BPS and RPA to become very similar, but with a significant bound effect missing, shown by our full model (solid). We also see the previously mentioned issue with the low-velocity form of CBC, which causes a large secondary peak at low velocity. At 100 times compressed, BPS and RPA disagree significantly with different temperature dependencies. 


\subsection{Application to Warm Dense Carbon}
Our last application will be to the one experiment that has measured stopping near the Bragg peak in the WDM regime\cite{Malko2022-ws}. This measurement showed a significant discrepancy with all existing stopping models, including TD-DFT, in that the overall stopping power was found to decrease as the sample was heated into the WDM regime. We consider whether an improved bound state treatment could be the cause of the reported deficit, and apply our model to test this.

In this experiment a carbon foil is laser heated to the WDM regime while a target normal sheath acceleration (TNSA) proton beam is fired with an incident time of approximately $100$ ps after the laser turn-on, though with a large proton pulse width of approximately $400$ ps. 

\begin{figure}
    \centering
    \includegraphics[width=1\linewidth]{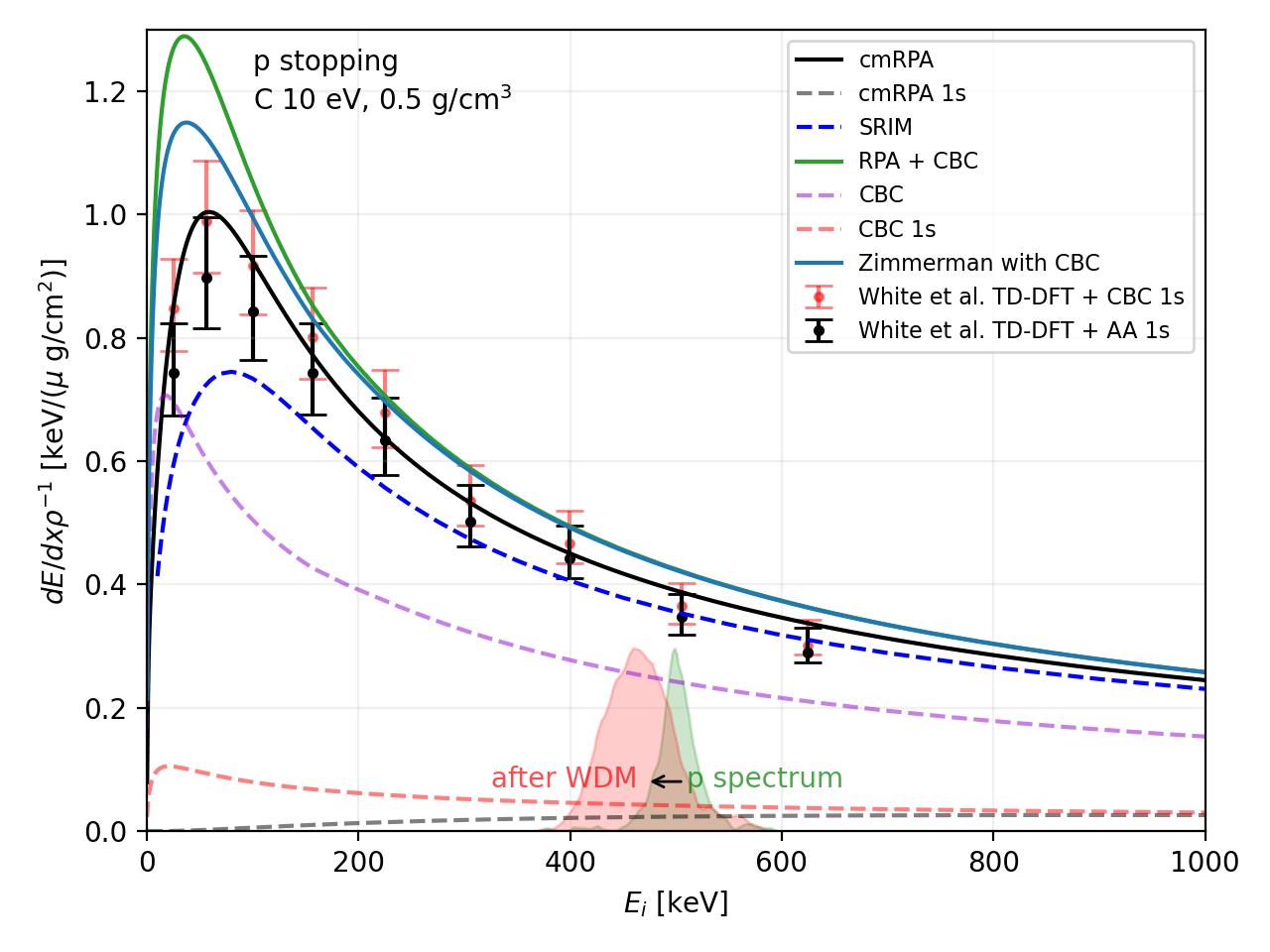}
    \caption{The fiducial point in stopping power parameter space used for model comparison in Malko et al. 2022 \cite{Malko2022-ws}. We show our prediction (black solid), our 1s contribution (grey dashed) vs the 1s CBC contribution  (red dashed), and the original (red errorbars) TD-DFT data \cite{White_2022} and the version with our 1s contribution replacing that of CBC (black errorbars). We also compare to the SRIM ambient conditions model (blue dashed), RPA with CBC bound contribution (green solid), and the Zimmerman RPA fit with CBC bound (blue solid), as well as the CBC by itself (purple dashed). The initial proton beam energy distribution is shown at the bottom in green, and the distribution after passing through the laser heated carbon is in red. }
    \label{fig:C_10eV_0.5gpercc}
\end{figure}
Stopping power models were bench-marked at the point $\rho=0.5$ g/cm$^3$ and $T=10$ eV, displayed here in Fig.~\ref{fig:C_10eV_0.5gpercc}. The TD-DFT data shown in red error bars is from White et al. \cite{White_2022} though it overlaps exactly with the same calculation in the Malko et al. paper, but contains more points. This data was computed using four valence electrons, and the two 1s electrons were treated with the CBC model \cite{Barriga-Carrasco_Casas_2013}, which we verify against \cite{White_2022} results. Since the CBC model overestimates stopping at lower velocities, we modify the data by replacing the CBC 1s contribution with our 1s contribution (black error bars). This shift is still within the TD-DFT error bars, but it is a physical overall shift that brings the TD-DFT stopping data lower. While this correction does push TD-DFT in the direction observed by experiment, the effect is largely insignificant at the $E_i = 500 {\rm keV} \implies v_i = 4.47 $ a.u. proton velocity probed.    

\begin{figure}
    \centering
    \includegraphics[width=1\linewidth]{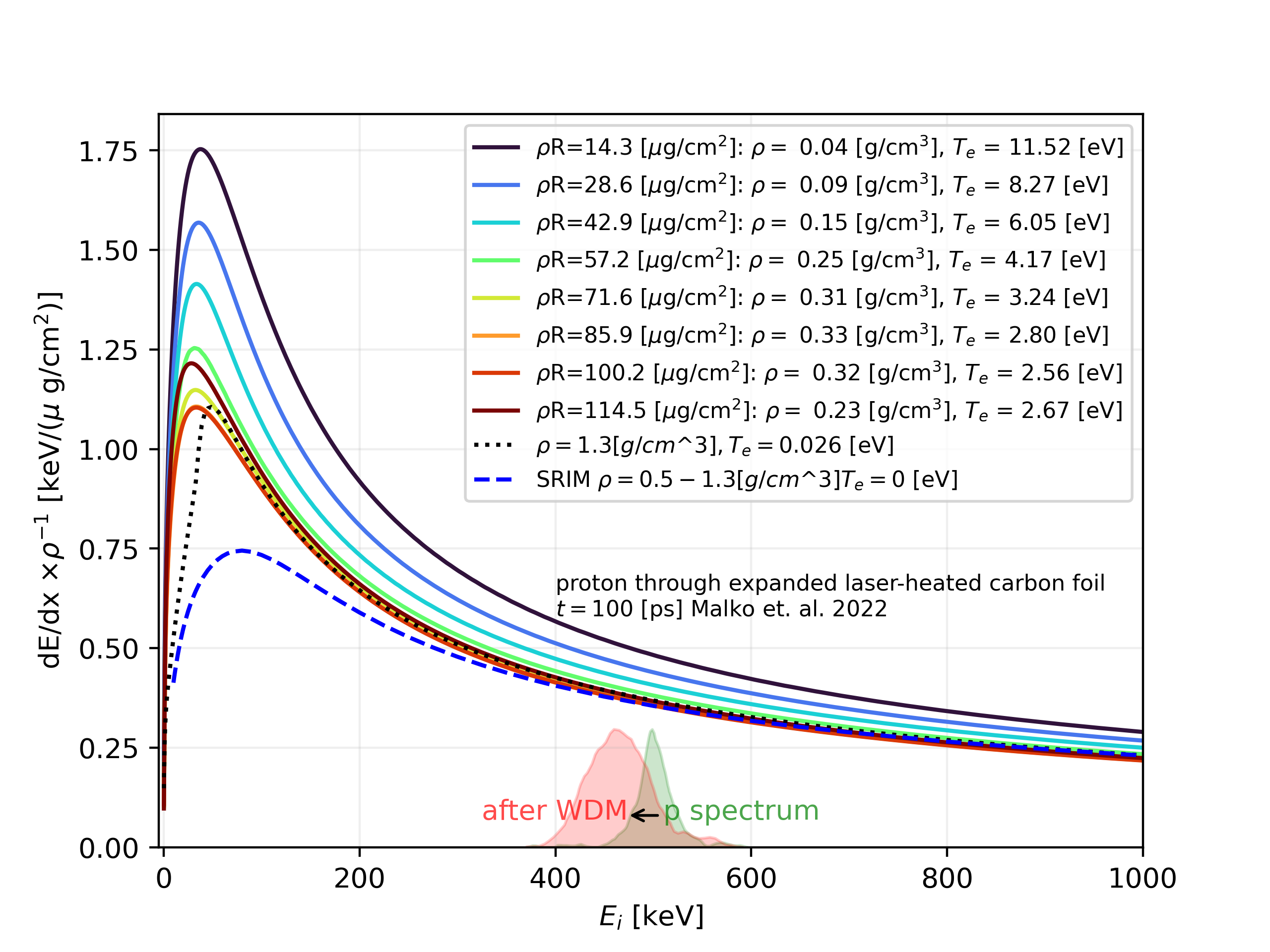}
    \caption{Stopping power for points evenly spaced in $\rho R$ along the proton beam line of sight at $100$ ps after the laser turns on for Malko et al. 2022 \cite{Malko2022-ws} evaluated with our model, and compared to SRIM data, which is invariant under both temperature and density for these conditions when divided by the density. The initial proton beam energy distribution is shown at the bottom in green, and the distribution after passing through the laser heated carbon is in red.}
    \label{fig:Malko_100ps_dE_dx}
\end{figure}

Hydrodynamic modeling done in the Malko et al. experiment reveals a large range of temperature and density conditions probed by the proton beam. Using this data, we compute stopping at fixed time $t=100$ ps after laser turn on, near the average time when the protons reach the carbon foil, shown in Fig.~\ref{fig:Malko_100ps_dE_dx}. The solid color lines denote stopping predictions uniformly spaced in $\rho x$ space from their hydrodynamic data\cite{Malko2022-ws}, compared to our ambient carbon foil (black dotted) and the SRIM model which is both temperature and density independent. We see the stopping power when divided by the density, is nearly condition independent except at the extreme edges of the expanded plasma where the density is lower and the temperature higher, where we see an increase in the stopping power. Additionally, the middle of the expanded foil where the density is highest is very similar in stopping to the conditions expected in the fiducial point Fig.~\ref{fig:C_10eV_0.5gpercc} and the SRIM data. From this plot it appears unlikely that more detailed bound state stopping is the cause of the observed deficit in stopping. 
\begin{figure*}[t]
    \centering
    \includegraphics[width=1\textwidth]{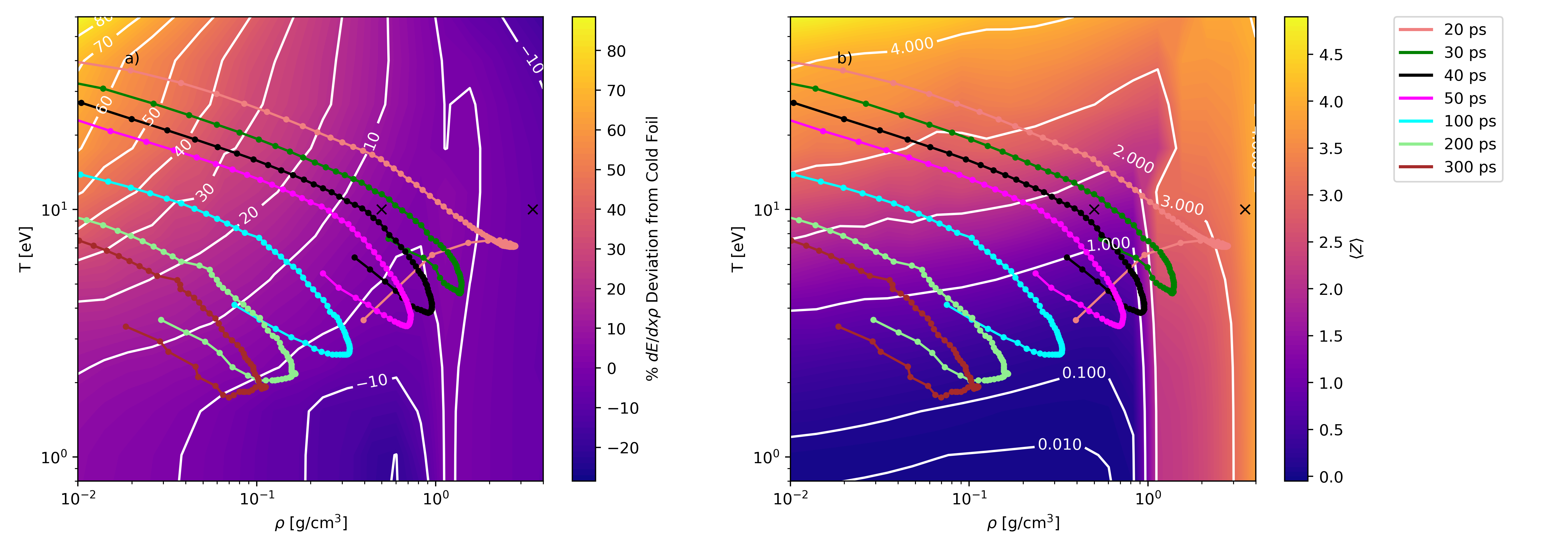}
    \caption{ {\bf a)} Deviation in stopping power for a 500 keV proton from our cmRPA model with respect to the cmRPA-derived stopping at solid carbon at 1.3 g/cm$^3$. Each colored line corresponds to the phase space trajectory a proton experiences at a given time point in the experiment after laser turn on, using hydrodynamic simulation data from \cite{Malko2022-ws}.   {\bf b)} The ionization state obtained from our AA input.   }
    \label{fig:Malko_contours}
\end{figure*}

\begin{table}[t]
    \centering
    \begin{tabular}{ccccl}\hline
 Method& subshell& $K$ [a.u.] & $\langle r^2 \rangle$ [a.u.]&$I$ [a.u.]\\\hline
         Hartree-Fock \cite{Barriga-Carrasco_Casas_2013}&  1s&   16.053 & 0.097 & 18.176\\
         &  2s&  1.546 &  3.038 &1.009\\
         &  2p&  1.230&   3.890 &0.795\\
         Average Atom&  1s&  15.74714  &      0.10056   &  17.69735\\
         &  2s&  1.94347&  2.58329&1.22664\\
         &  2p&  1.58946&  2.98496&1.03198\\
    \end{tabular}
    \caption{The parameters necessary as input to the CBC model, with Hartree-Fock isolated atom data from \cite{Barriga-Carrasco_Casas_2013}  and alternatively evaluated with our finite temperature and density AA orbitals. This data is for carbon at the conditions in Fig.~\ref{fig:C_10eV_0.5gpercc}.}
    \label{tab:Malko_CBC}
\end{table}

To further this idea, we compute stopping over nearly the full range of temperature and density conditions observed in the experiment over all times. Since the proton is at approximately 500 keV throughout the stopping, we assume this is fixed, and compute the percent difference in stopping power at every point from what we predict at ambient conditions, shown in the left panel of Fig.~\ref{fig:Malko_contours}. We see at high temperatures and densities the stopping is higher due to the larger ionization, and there is a small region of non-trivial shape where there is a stopping deficit for higher densities and lower temperatures. In this space we plot hydrodynamic data from \cite{Malko2022-ws} as trajectories through this space, shown by the various colored lines from 20 ps to 300 ps after laser turn on. On the right panel we show the corresponding $\Zbar$. The space between each dot on these trajectories is of equal $\delta x \rho$ so denser places correspond to larger regions of the expanded carbon foil. 

\begin{figure}
    \centering
    \includegraphics[width=1\linewidth]{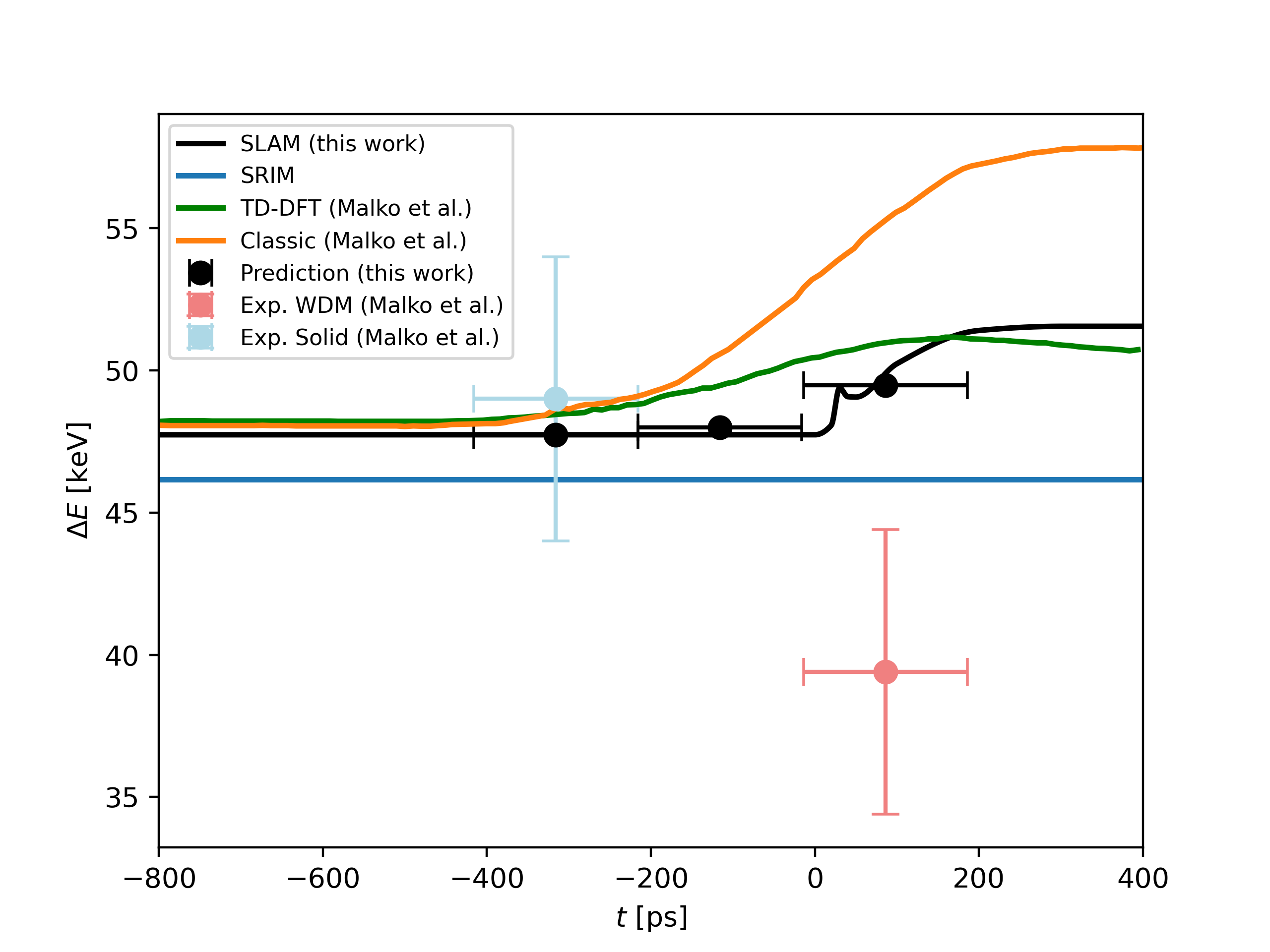}
    \caption{We show our time dependent energy loss prediction from our model (black solid) integrated over the three proton beam bunches (black errorbars) compared to data from \cite{Malko2022-ws} for the classical models (orange), TD-DFT(green), and data (light blue and light red errorbars), as well as the ambient SRIM expectation (blue solid). }
    \label{fig:Malko_final}
\end{figure}

Based on Fig.~\ref{fig:Malko_contours} the proton has a higher stopping power in much of the parameter space the beam probes and only in a relatively small region in the middle of the expanded foil is there a region with a slight dipping in stopping, at least according to the hydrodynamic data and our model. We then compute a time dependent energy loss to see the net effect, shown as the purple line in Fig.~\ref{fig:Malko_final}. Then integrating the three proton beam shot temporal distributions in \cite{Malko2022-ws} over this trajectory sampling, we obtain three stopping predictions which again qualitatively predict a rise in stopping, largely due to the time spent in the hot low density regime of the expanded carbon where stopping is high. 

Thus, without substantially altered hydrodynamic data or other inputs we also predict that this integrated stopping measurement should show a rise in stopping of the heated carbon foil. We will however note that from looking at the right panel of Fig.~\ref{fig:Malko_contours}, that a significant portion of the parameter space that the experiment probes is in a low density, low temperature regime where our AA model predicts low ionization; if significant bonding is occurring there, then it is possible that the spherical symmetry assumption implicit in our AA model is violated significantly. Additionally, there is no TD-DFT data to compare with in this regime, leaving the possibility that there are yet unknown and non-trivial stopping effects on the protons.

\section{Conclusions}
We developed an atomic orbital-based method for computing the energy loss function and particle stopping power using the average atom that mixes bound electron response with the Lindhard continuum response. Despite the computational efficiency of this method, we show very good agreement with near-ambient condition experimental data and TD-DFT simulations when restricted to the region above the Bragg peak where linear response is a good approximation. We then compute stopping power for protons and alphas over a wide range of conditions in the warm dense matter and high energy density regimes.

In addition, we make several new physical insights. We find channel mixing via RPA between bound and free transitions leads to non-trivial bound-free corrections that can deviate from a simpler stopping power model where responses are only linearly added. We also predict a non-zero bound-bound excitation stopping contribution at low velocities in plasmas with partial occupation of bound states. We  then show a non-zero bound component to stopping for high energy density tungsten relevant for inertial confinement fusion. Lastly, we applied our model to the warm dense matter, partially ionized carbon experiment by Malko et al. We find good agreement with the time dependent density functional simulations, and demonstrate that within the constraints of the model that improved bound state modeling is unlikely to be the source of the experimental discrepancy with theory, suggesting more subtle physical or experimental effects are at play. 

Further work is needed to improve the treatment of the free-free response, which is treated at an inconsistent level with the bound transitions, resulting in a small all-electron f-sum rule violation. Additionally, effects such as hard binary scattering, charge capture, electronic correlations, and plasmon scattering-induced broadening are missing from the current implementation. Consistently implementing these corrections is subject to future work.

This work establishes a framework for combining bound and free electronic stopping that balances computational efficiency with physical accuracy, enabling robust predictions across the wide parameter space relevant to inertial confinement fusion experiments, ion beam experiments, and other applications where stopping on bound states needs to be considered.

\section{Acknowledgments}
The authors would like to thank Alexander White and Alina Kononov for many useful discussions, and Jackson White and Hoang Bao Tran Tan for advice on the matrix element calculations. 

Research presented in this article was supported by the Laboratory Directed Research and Development program of Los Alamos National Laboratory under project number 20250614CR-NLS. This research used resources provided by the Los Alamos National Laboratory Institutional Computing Program. Los Alamos National Laboratory is managed by Triad National Security, LLC, for the National Nuclear Security Administration of the U.S. Department of Energy under contract 89233218CNA000001. The U.S. Government retains an irrevocable, nonexclusive, royalty-free license to publish, translate, reproduce, use, or dispose of the published form of the work and to authorize others to do the
same for U.S. Government purposes.

\section*{Author Declarations}
The authors have no conflicts to disclose.

\section*{Data Availability}
    The data that support the findings of this study are available from the corresponding author upon reasonable request.
\bibliography{bib.bib}

\appendix

\section{Free-Free With Modified Density of States} \label{app:Lindhard_improved_DOS}
As mentioned throughout the article, our treatment of the free-free states is inconsistent with the bound state treatment in that we use Lindhard, which approximates an AA based f-f treatment by assuming an ideal density of states, as well as matrix elements based on plane-waves. In the low-k limit this is especially problematic since it is biased towards low-k continuum states which are very far from plane waves. Correcting this is difficult. However, one can expect to get part of the way to an improved free-free treatment via improving the DOS. 

For this we explore the method presented in \cite{hentschel2023improving} in which the ideal DOS in the Lindhard integral is replaced by the AA DOS, 
\begin{align}
    \Pi^{ff}_{aa}(k, w) = \int \frac{d^3 \p}{(2\pi)^3} \frac{g_{\rm DOS}^{AA}(E_p)}{g_{\rm DOS}^{ideal}(E_p)} \frac{f_a(\p) - f_a(\p + \k) }{\omega + E_a(\p) - E_a(\p + \k) + i\Gamma}. \label{eq:mod_lindhard}
\end{align}
There is only a single integral over continuum momenta here, as opposed to the double sum in Eq.~\eqref{eq:chi_0} because the Lindhard matrix elements contain plane waves that collapse one integral. We note that one can prove analytically and numerically verify that this construction satisfies the f-sum rule for $\Zbar$ electrons, by reverting to the double integral form over both initial and final states.  

\begin{figure}[t]
    \centering
    \includegraphics[width=1\linewidth]{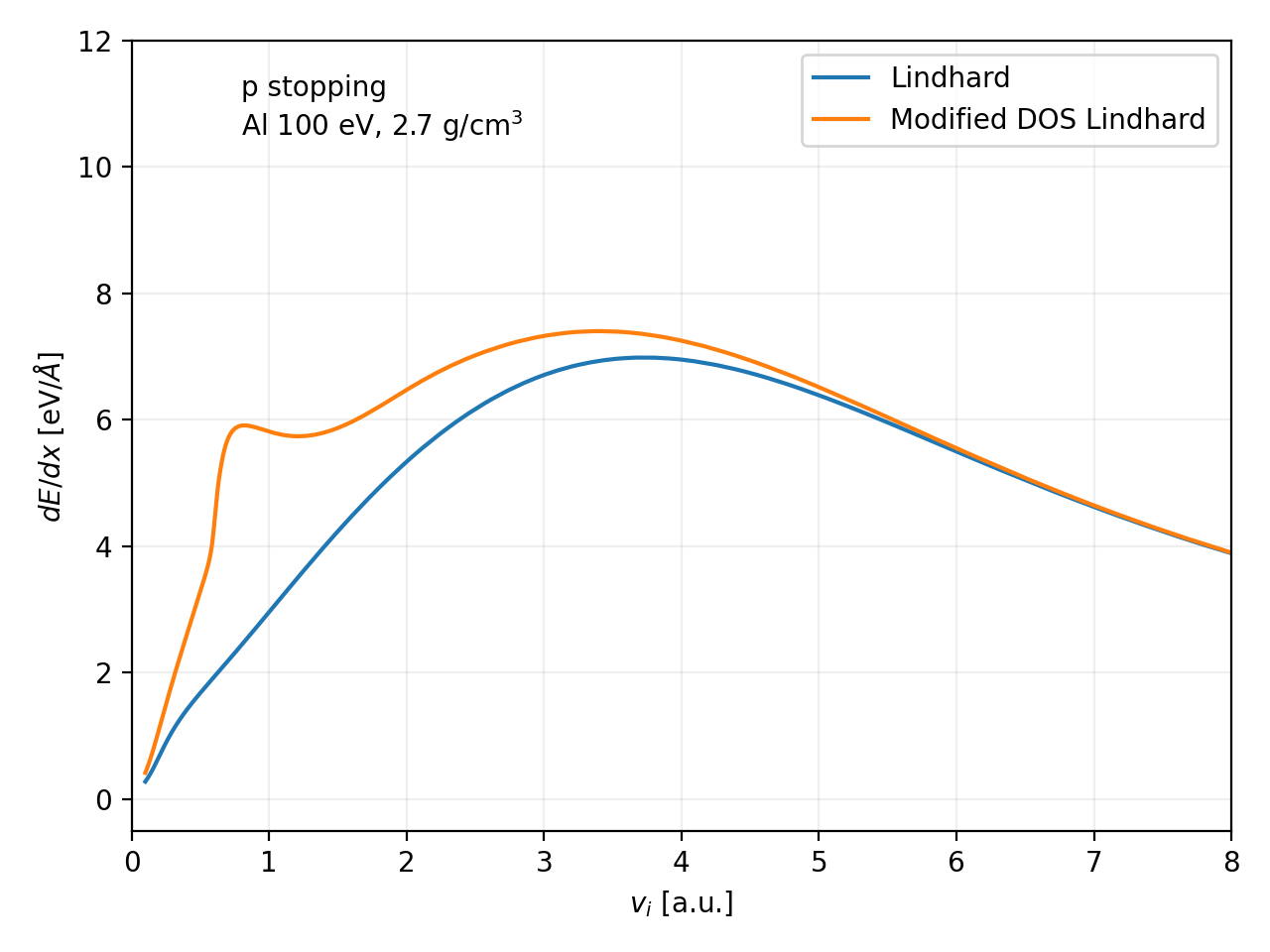}
    \label{fig:SP_AADOS}
    \caption{For Al at $T=100$ eV and $\rho=2.7$ g/cm$^3$, we show the stopping power using Eq.~\eqref{eq:mod_lindhard} (orange) compared with standard Lindhard (blue). }
\end{figure}

This correction is small if the density of states is similar enough to a free electron gas, however since the matrix elements remain to be that of plane wave transitions, it is not guaranteed to improve matters. In Fig.~\ref{fig:SP_AADOS} we show this construction for the case of Aluminum at a temperature of 100 eV where the AA model predicts a sharp resonance in the continuum corresponding to a quasibound 3d state. The impact of this on the stopping is shown in panel {\bf b)}  where there is an unphysical peak at low velocity. Here this peak is due to the large difference between the quasibound orbital and the Lindhard plane waves, which have very different transition rates.     

\section{Restriction to Ion-Sphere Radius for Stopping Power}\label{app:truncated}
The above results assume a sufficiently large extended radius such that the entire bound state is captured, which is required for orthogonality between states, and for the sum rules above to hold. In our AA model, this sphere is extended by solving for the AA orbitals assuming a potential of zero outside the ion-sphere radius out to a sufficiently large radius to ensure asymptotic convergence of the sum rules. Instead, one can cutoff integrals at the ion-sphere radius, giving up orthogonality and thus exact sum rules, but generally having a very limited effect on the total stopping power at the gain of many times faster computation.   

In this paper we choose the restricted model, and regularly check that the impact of this approximation is within the accuracy that we expect this model to be accurate anyway. One final piece that is required to obtain decent results is the modification of Eq.~\eqref{eq:chi_bf} to replace specifically the $j_0(kr)$ component with $j_0(kr) - 1$. When bound states are numerically orthogonal, this modification does nothing, but when there is a small non-orthogonality due to the finite sphere, in the $k\to0$ limit, this correction ensures the overlap integrals of Eq.~\eqref{eq:chi_bf} correctly vanish as $k^2$.

\begin{figure*}[t]
    \centering
    \begin{minipage}{0.48\textwidth}
        \includegraphics[width=1\textwidth]{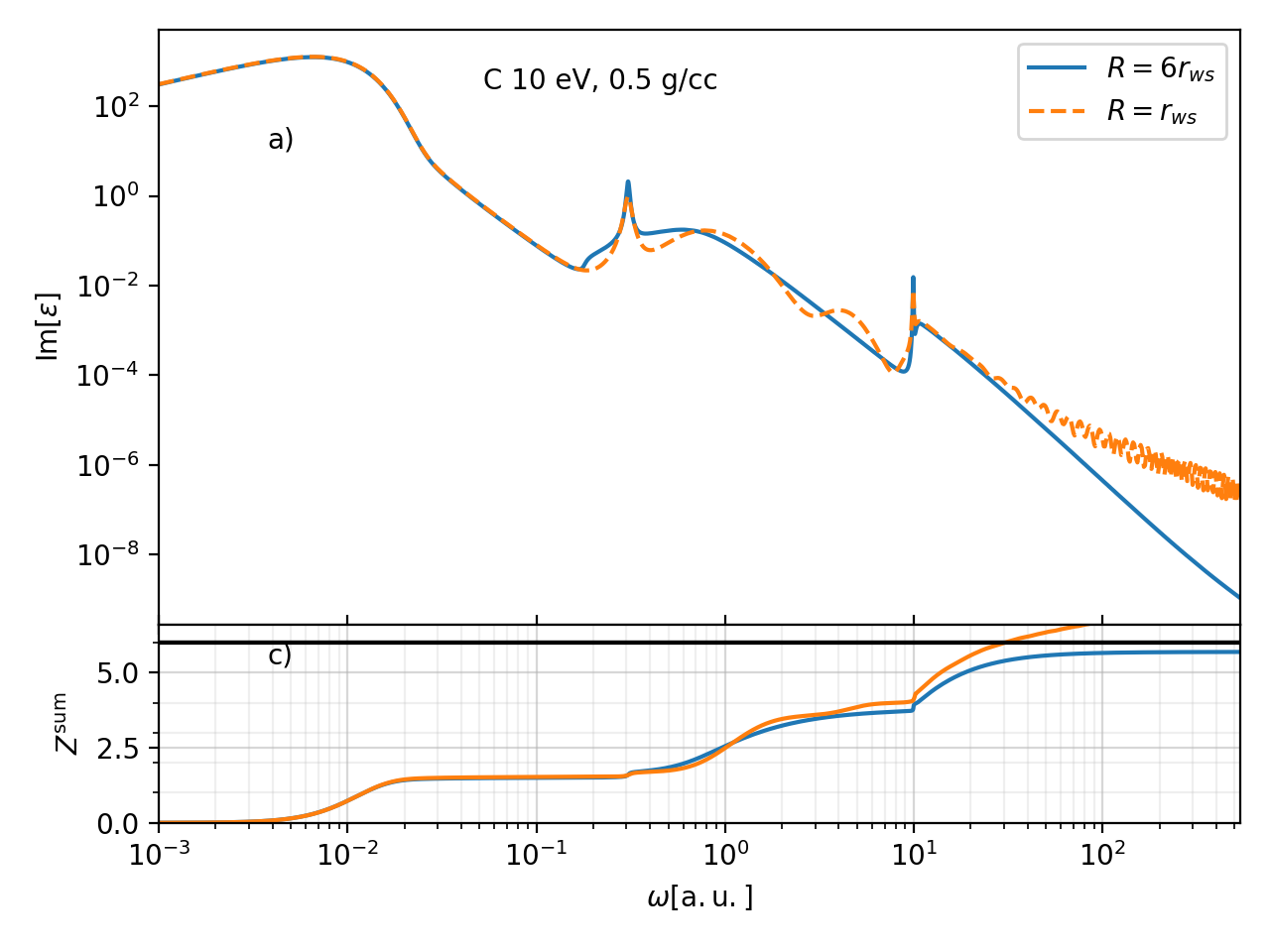}
    \end{minipage}%
    \begin{minipage}{0.48\textwidth}
        \centering
        \includegraphics[width=1.0\textwidth]{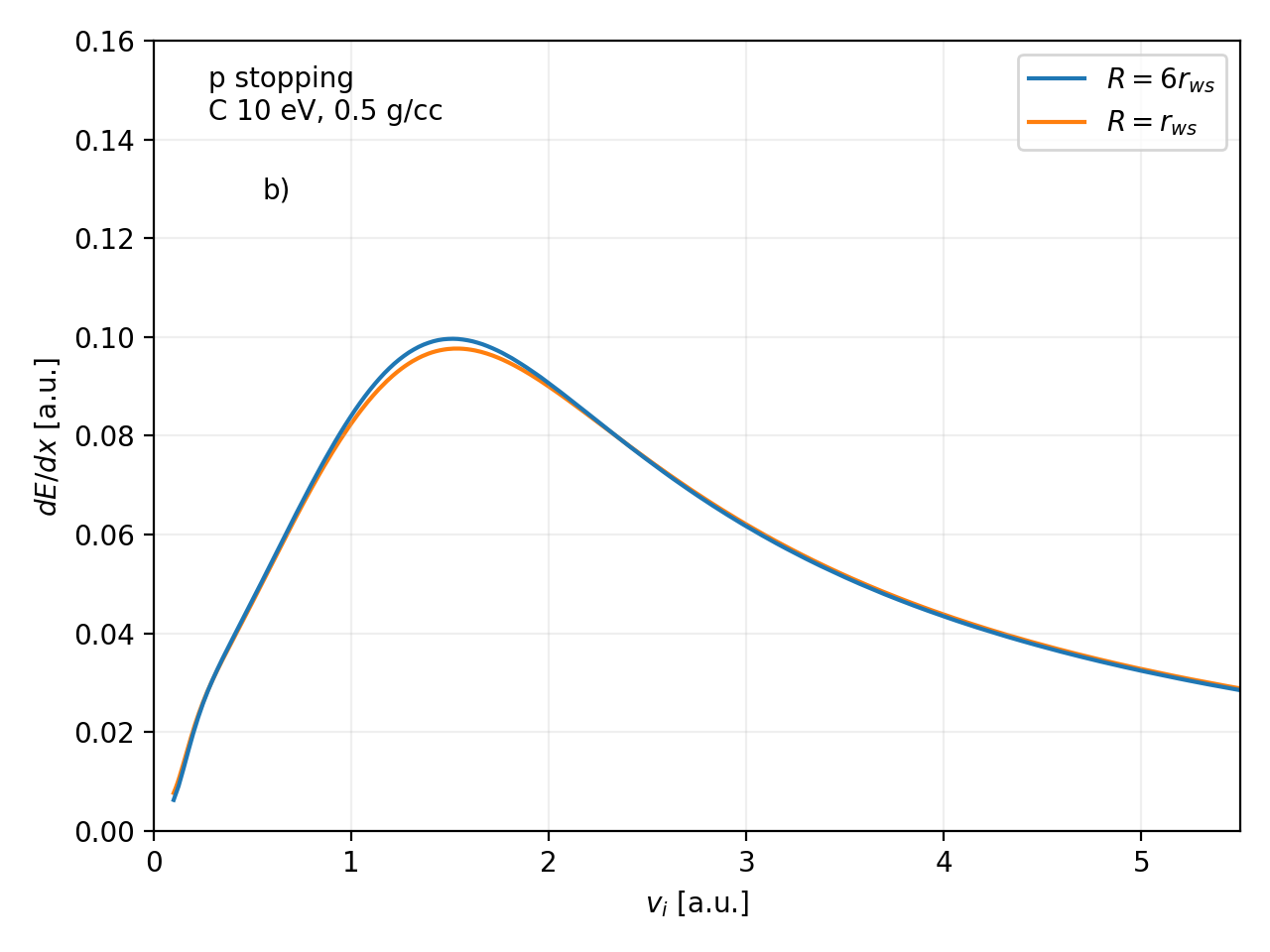} \\ 
    \end{minipage}
    \caption{ We show carbon at $T=10$ eV and $\rho=0.5$ g/cm$^3$ where our AA model predicts bound states at energies -10.1 Ha (1s), -0.48 Ha (2s), and -0.18 Ha (2p) with $\Zbar=1.51$. In {\bf a)} is the imaginary part of the dielectric for the model with extended maximum radius (blue), and the one where all integrals are restricted to the ion-sphere radius (orange dashed), violating orthogonality for the 2s, 2p states which are not entirely contained in the ion sphere radius. The bottom shows sum rule Eq.~\eqref{eq:sum_diel_1} normalized to recover the number of electrons, $Z$. In  {\bf b)} we show the corresponding effect on stopping power and it is only visible at the Bragg peak where our linear response model is already an approximation.}
    \label{fig:Rmax_test}
\end{figure*}

In Fig.~\ref{fig:Rmax_test} we can see the impact of this approximation the dielectric in the left panel, and the stopping on the right. The lack of perfect orthogonality between bound and free states in the restricted sphere shows up as oscillations in frequency as we move up in continuum energy and encounter free states with varying missing portions. In terms of the impact on the f-sum rules, lower panel on the left, the effect is modest except at high frequencies where the behaviour diverges. The small violation of the f-sum rule in the extended sphere (blue) is due to the use of Lindhard for the f-f component.

\begin{figure*}[t]
    \centering
    \begin{minipage}{0.48\textwidth}
        \includegraphics[width=1\textwidth]{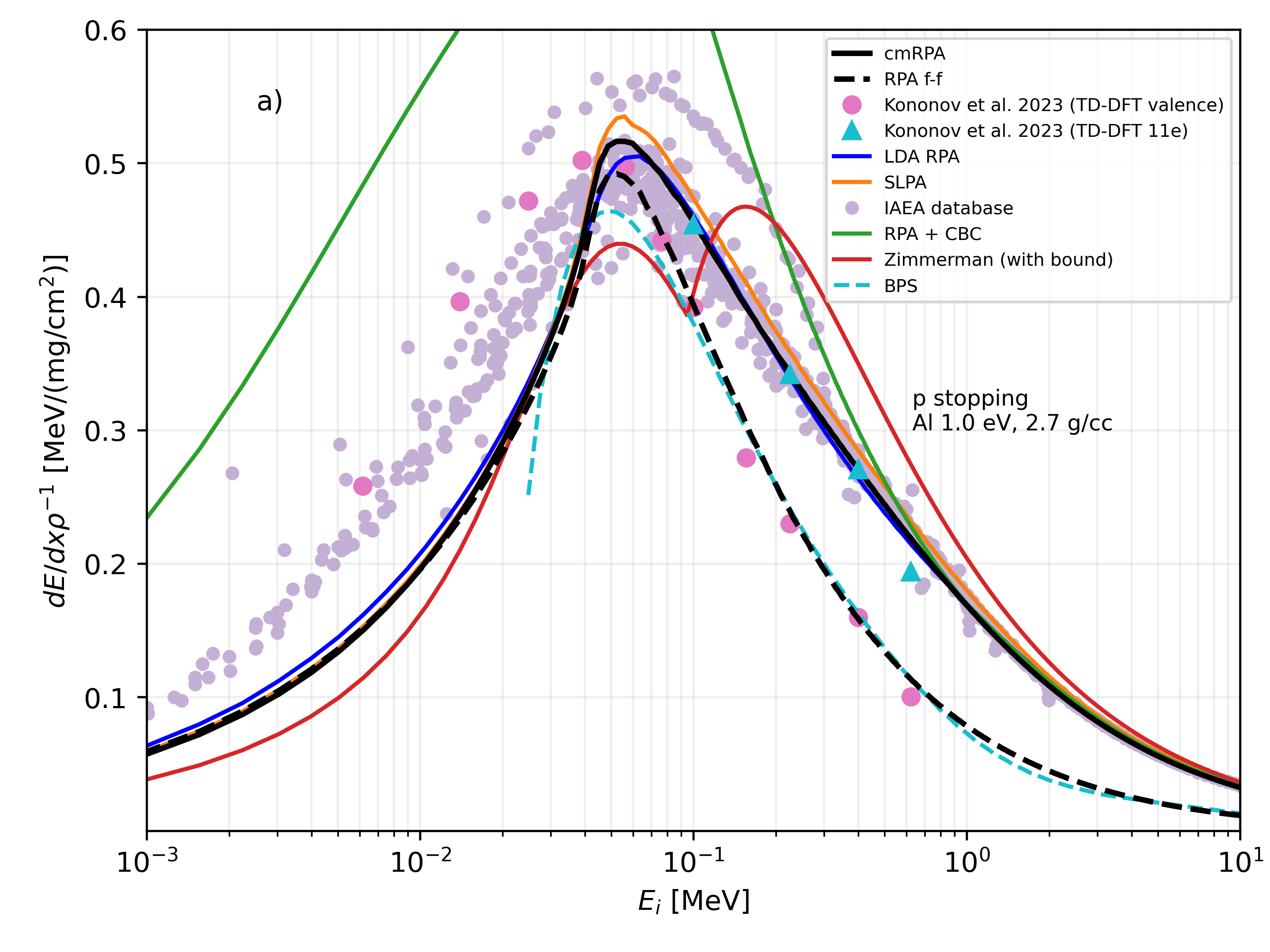}
        \includegraphics[width=1\textwidth]{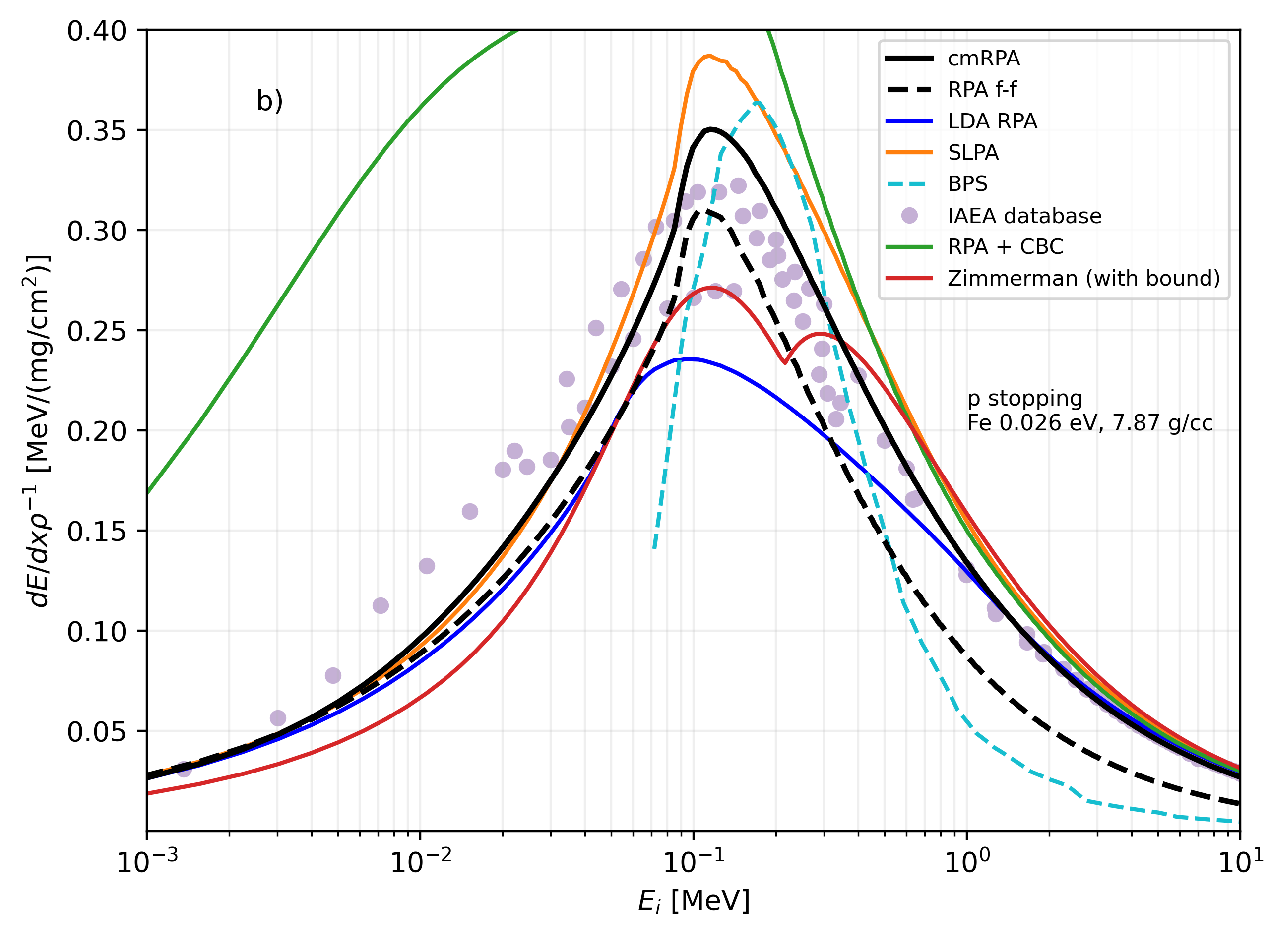}
    \end{minipage}%
    \begin{minipage}{0.48\textwidth}
        \centering
        \includegraphics[width=1.0\textwidth]{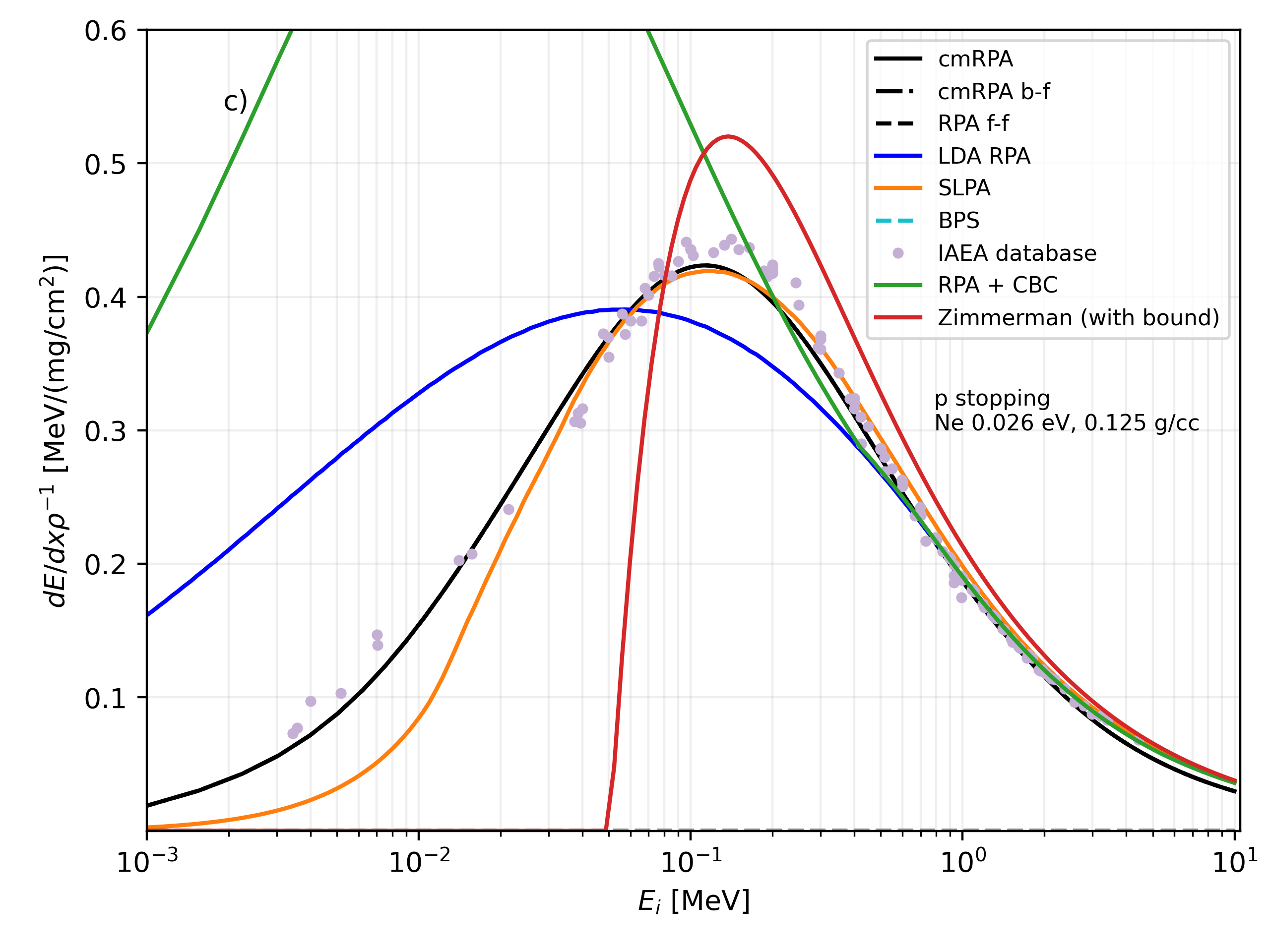}
        \includegraphics[width=1.0\textwidth]{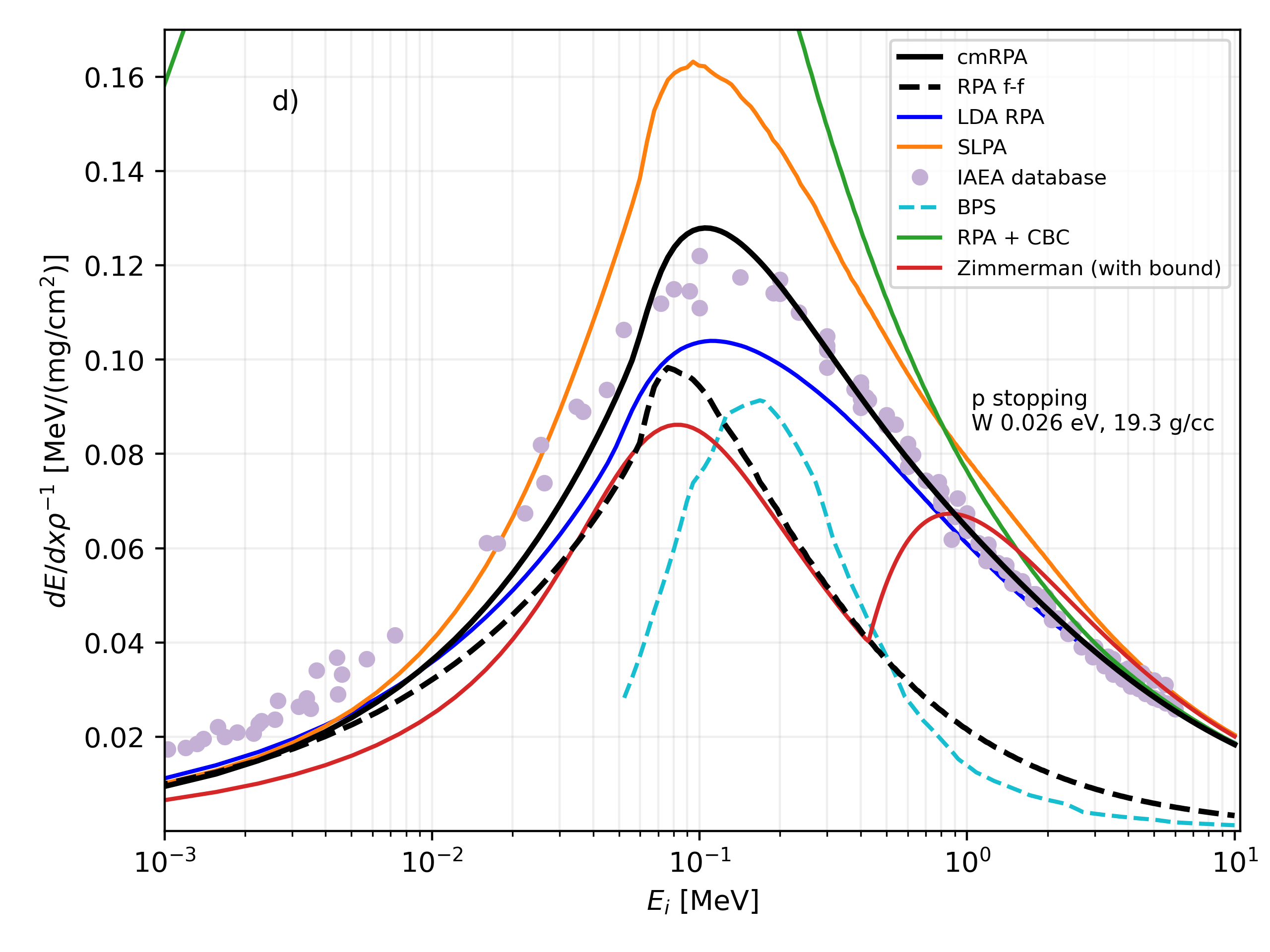}
    \end{minipage}
    \caption{In a more detailed version of Fig.~\ref{fig:ambient_benchmark}, we plot proton stopping near ambient condition for the IAEA experimental database (purple circles) \cite{IAEA}, the Zimmerman model \cite{zimmerman1990recent} (red solid), the CBC model for bound stopping with RPA for free-free (green solid), the local density approximation \cite{wang1998unified, faussurier2010equation} (blue solid), the bound state resolved SLPA \cite{PhysRevA.66.042902}, the BPS model \cite{BROWN2005237} against our stopping model prediction (black solid), the f-f contribution (black dashed), the b-f (black dash-dot) in target materials {\bf a)} aluminum at ambient conditions $\rho=2.7$ g/cm$^3$, $T=1$ eV, $\Zbar=3.0$ where we also have TD-DFT data \cite{Kononov2023-wl} for a 3e pseudopotential (pink circles) and an 11e pseudopotential (turquoise triangles), {\bf b)} iron at ambient density, 7.87 g/cm$^3$, 0.026 eV, $\Zbar=8.0$ {\bf c)} neon gas at 0.125 g/cm$^3$, 0.026 eV, and $\Zbar=0$,  {\bf d)} tungsten at ambient conditions, $T=0.026$ eV, $\rho=19.3$ g/cm$^3$ and $\Zbar=6.0$ . 
    }
    \label{fig:ambient_benchmark_model_comparison}
\end{figure*}


\section{Ambient Model Comparison}\label{app:extra_ambient}

In this appendix we compare ambient stopping model data for several models. In particular, in Fig.~\ref{fig:ambient_benchmark_model_comparison} we show a more detailed version of Fig.~\ref{fig:ambient_benchmark} which compares to a variety of other models in the literature that we implement. These require our AA model output including the ionization, binding energies and orbital kinetic energies. The Zimmerman model \cite{zimmerman1990recent} (red solid) is largely a correction to a fit of RPA by Maynard and Deutsch \cite{maynard1985born} but also adds a Bethe-like mean-excitation energy for the bound-free contribution. We correct it slightly by flooring the coulomb log at zero. While this correction is reasonable at high proton energies, the abrupt turn-on of the bound contribution gives a nonphysical second local maximum. The CBC \cite{Barriga-Carrasco_Casas_2013} (solid green) model accounts for bound states individually with Bethe-like terms based on average atom or Hartree-Fock output so the high energy limit is generally quite close and asymptotically converges to experimental results. The low velocity, however, contains a second maximum independent of the Bragg peak; at near ambient conditions this manifests as an overestimate of the low velocity stopping, but at higher densities and temperatures the second peak becomes distinct from the Bragg peak. We also see the BPS\cite{BROWN2005237} model does poorly in these conditions since it is a low density expansion, excepting the aluminum case whose valence electrons behave nearly like a free electron gas; we expect its accuracy to be much improved in hotter conditions.

The other two models shown, LDA and SLPA, have been used extensively in near ambient conditions \cite{wang1998unified,MONTANARI2024165336,barriga2022stopping, PhysRevA.96.012707, PhysRevA.105.062814, PhysRevA.66.042902} but generally only the LDA, and particularly the SCAALP version of it\cite{faussurier2010equation} has seen use in plasma physics. The idea is to treat the bound state response as a collection of free states at density conditions corresponding to the electron density output from the electronic structure model used, here the AA. The response assumed for each density is the finite-temperature Lindhard, integrated over the whole atom. SLPA takes this a step further and essentially does this per bound state density, with an additional ionization cutoff energy. By using Lindhard to model bound states, these models are physically very approximate, but conserve important sum rules that allow for reasonable answers to be obtained everywhere. Recent uses of SLPA \cite{PhysRevA.105.062814} actually incorporate additional physics such as strong binary collisions and the Mermin improvement to obtain better results, but we don't consider that here. Without these modifications, we find the agreement of these models with experimental data and simulation to be worse across the cases considered. We also note that though the theory is simpler, it requires computing finite-temperature Lindhard across $k,\omega $ space for a large range of electron densities. For similar parallelization methods we find the cost of LDA to be similar to our model, and SLPA to be more expensive.

\section{Local Field Corrections} \label{app:LFC}

The above RPA resummation is the exact result in the limit where the exchange-correlation potential is zero. Including the effect of exchange and correlation is typically done in terms of a local field correction. The local field correction for the full response function including all states is unknown and would for example, involve solving heterogeneous linear response TD-DFT with some adiabatic approximation to the exchange-correlation functional as in \cite{gill2021time}. 

However, corrections to the free electron gas are well studied \cite{Faussurier2025, dornheim2021analytical} and we choose to implement the analytical static LFC in \cite{dornheim2021analytical} which is available by that group as a python package. We suggest correcting only the free portion, as 
\begin{align}
    \Im[\epsilon^{-1}] &= \frac{V_k \Im[\chi^0_{bb}+ \chi^0_{bf} + \chi^0_{ff}]}{ |1 - V_k(\chi^0_{bb}+ \chi^0_{bf} + \chi^0_{ff}(1-G_{ff}))|^2 }. \label{eq:mixed_inv_eps_LFC}
\end{align}

\begin{figure}
    \centering
    \includegraphics[width=1\linewidth]{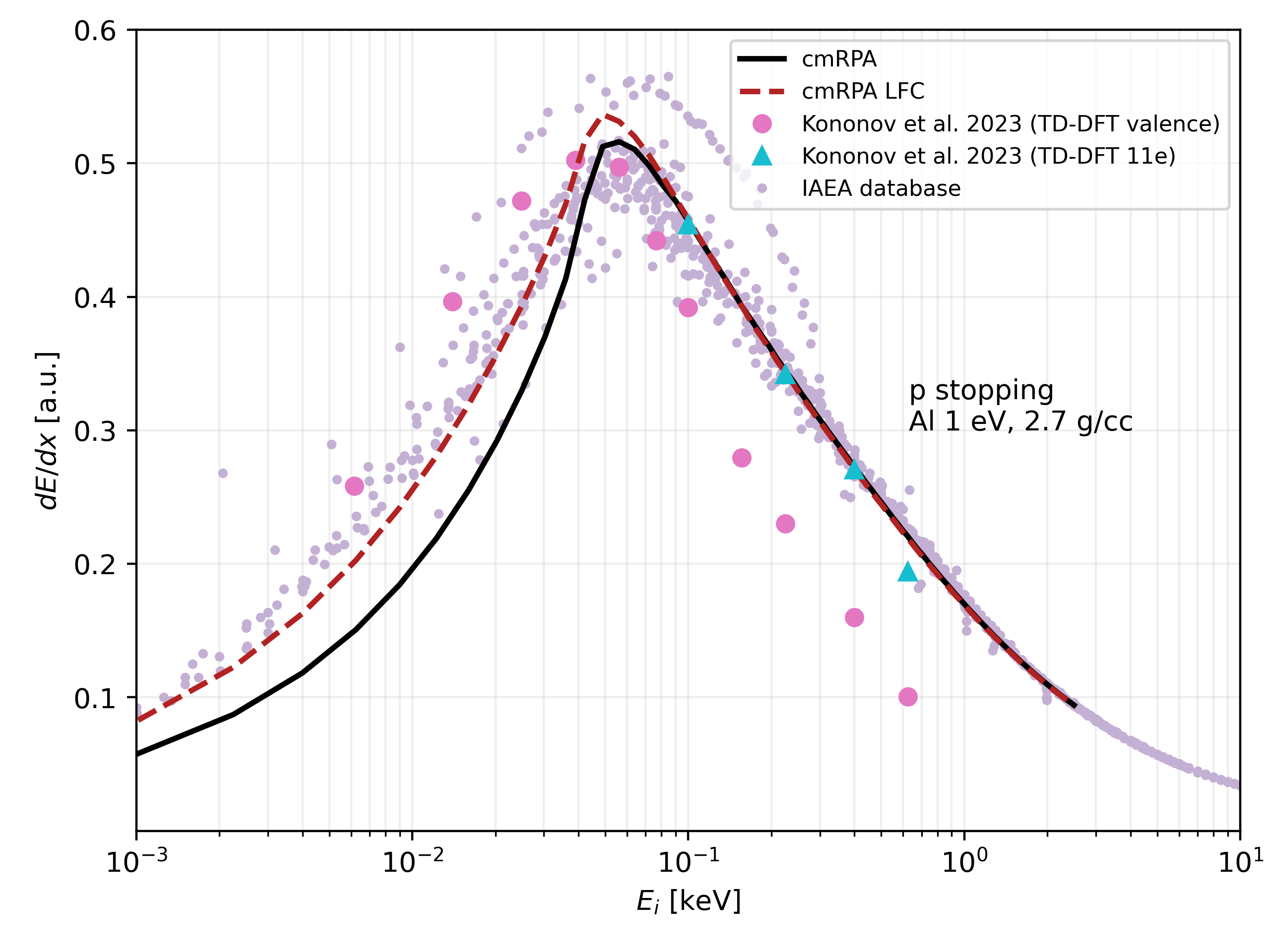}
    \caption{We show again the result in Fig.~\ref{fig:ambient_benchmark} {\bf a)}, but with the LFC in \cite{dornheim2021analytical} applied as in Eq.~\eqref{eq:mixed_inv_eps_LFC}, shown now in red dashed versus without the LFC in solid black.}
    \label{fig:Al_LFC}
\end{figure}
We tested the impact of this improvement on the case of Aluminum from Fig.~\ref{fig:ambient_benchmark}, which we show in Fig.~\ref{fig:Al_LFC}. We can see the stopping power with the LFC (red dashed) increases the low velocity stopping, as in \cite{Faussurier2025} bringing it closer to experimental data. It also tends to increase the Bragg peak when it is already typically too high compared to data and TD-DFT simulations, so whether it is considered a model improvement to include a LFC is application dependent. We note this LFC will not affect low-k sum rules, nor stopping in high temperature conditions. 

\end{document}